\RequirePackage{fix-cm}
\documentclass[twocolumn,epjc3]{svjour3}
\smartqed
\RequirePackage{graphicx}
\RequirePackage{mathptmx}
\RequirePackage{amsmath}
\RequirePackage{amsfonts}
\RequirePackage{amssymb}
\RequirePackage{latexsym}
\RequirePackage{bm}
\RequirePackage{widetext}
\RequirePackage{upgreek}

\usepackage{bm}

\newcommand{\be}{\begin{equation}}
\newcommand{\ee}{\end{equation}}
\newcommand{\bea}{\begin{eqnarray}}
\newcommand{\eea}{\end{eqnarray}}

\journalname{Eur. Phys. J. C}
\begin{document}

\title{Growth of matter overdensities in non-minimal torsion-matter coupling theories}

\author{Manuel Gonzalez-Espinoza\thanksref{e1,addr1}
    \and
		Giovanni Otalora\thanksref{e2,addr1}
	  \and
		Joel Saavedra \thanksref{e3,addr1}
  \and
	Nelson Videla \thanksref{e4,addr1}}

\thankstext{e1}{e-mail: manuel.gonzalez@pucv.cl}
\thankstext{e2}{e-mail: giovanni.otalora@pucv.cl}
\thankstext{e3}{e-mail: joel.saavedra@pucv.cl}
\thankstext{e4}{e-mail: nelson.videla@pucv.cl}

\institute{Instituto de F\'{\i}sica, Pontificia Universidad Cat\'olica de
Valpara\'{\i}so, Casilla 4950, Valpara\'{\i}so, Chile \label{addr1}}

\date{Received: date / Accepted: date}

\maketitle

\sloppy

\begin{abstract}
We study the evolution of cosmological perturbations around a homogeneous and isotropic background in the framework of the non-minimal torsion-matter coupling extension of $f(T)$ gravity. We are concerned with the effects of the non-minimal coupling term on the growth of matter overdensities. Under the quasi-static approximation, we derive the equation which governs the evolution of matter density perturbations, and it is shown that the effective gravitational coupling 'constant' acquires an additional contribution due to the non-minimal matter-torsion coupling term. In this way, this result generalizes those previously obtained for the growth of matter overdensities in the case of minimal $f(T)$ gravity. In order to get a feeling of our results we apply them to the important case of a power-law coupling function, which we assume to be the responsible for the late-time accelerated expansion in the dark energy regime. Thereby,  analytic solutions for the matter density perturbation equation in the regime of dark matter dominance and the dark energy epoch are obtained, along with a complete numerical integration of this equation. In particular, we show that this model predicts a growth index larger than those obtained for $\Lambda$CDM model, indicating therefore a smaller growth rate. Concomitantly, we show that the model at hand is potentially capable in alleviating the existing $\sigma_{8}$-tension, being that it can provide us a $f\sigma_{8}$ prediction which is $\sim 4-5$ per cent below the respective prediction of concordance model.
\end{abstract}

\section{Introduction}\label{Introduction}

Dark energy is one of the most amazing findings in modern cosmology \cite{Riess:1998cb,Perlmutter:1998np,Ade:2013zuv,Aghanim:2018eyx}. This dark component is responsible by the accelerated expansion of the Universe and its nature is still one of the great mysteries of the Big Bang \cite{Copeland:2006wr,Frieman:2008sn1,Nojiri:2017ncd}. Furthermore, dark energy together with dark matter, another mysterious component \cite{Zwicky:1933gu}, constitute $68 \%$ and $27 \%$, respectively, of the total mass-energy of the present Universe, remaining only $5 \%$ for the normal baryonic matter \cite{AmendolaTsujikawa,FR-reviews7}. There are two principal ways to proceed in the study of the nature and properties of this dark energy entity. The first one is considering it as a new modified matter source which is described for example through a scalar field such as quintessence \cite{Ratra:1987rm,Copeland:1997et,Caldwell:1997ii,Barreiro:1999zs}, tachyon field \cite{Sen:2002nu,Sen:2002in,Padmanabhan:2002cp,Abramo:2003cp,Copeland:2004hq}, k-essence \cite{Chiba:1999ka,ArmendarizPicon:2000dh,ArmendarizPicon:2000ah},  or, dilatonic ghost condensate \cite{Piazza:2004df,Gasperini:2001pc,Szydlowski:2006pz}, etc, being that in all these models, the scalar field contributes with a negative pressure which drives the accelerated expansion. On the other hand, the second one alternative is represented by modified gravity theories, which are mainly based on quantum corrections to the Einstein-Hilbert action of General Relativity (GR), in the form of additional higher curvature terms such as $f(R)$ gravity theories \cite{FR-reviews1,FR-reviews2,FR-reviews3,FR-reviews4,FR-reviews5,FR-reviews6}. In this last approach, one may identify in the modified Friedmann equations an effective dark energy density and its corresponding pressure density, which comes to have an origin in quantum corrections to GR, and therefore it becomes conceptually different from a modified matter model \cite{AmendolaTsujikawa,FR-reviews7}.

It is well known that gravity can be described in terms of curvature, as is usually done in GR and $f(R)$ gravity theories, or through torsion, in which case we have the so-called teleparallel equivalent of GR or simply Teleparallel Gravity (TG) \cite{Einstein,TranslationEinstein,Early-papers1,Early-papers2,Early-papers3,Early-papers4,Early-papers5,Early-papers6}.  In TG the dynamical variables are the tetrad fields instead of the metric tensor $g_{\mu \nu}$, and the usual torsionless Levi-Civita connection of GR is replaced by the Weitzenb\"{o}ck connection, which has torsion but no curvature \cite{Aldrovandi-Pereira-book,JGPereira2,AndradeGuillenPereira-00}. So, TG is a classical gauge theory for gravitation based in the translation group, that due to existence of "soldering" between the Minkowski tangent space (fiber) and the spacetime (base space), it becomes a non-standard gauge theory, keeping nevertheless a remarkable similarity to electromagnetism, also a gauge theory for an abelian group \cite{Aldrovandi-Pereira-book,Arcos:2005ec}. It is worth noting that, the Lagrangian density of GR, the scalar curvature $R$, and the Lagrangian density of TG, the scalar torsion $T$, only differ in a total derivative term, and despite being conceptually speaking different theories they are equivalent in the level of field equations \cite{Aldrovandi-Pereira-book}. 

In the context of modified gravity theories one may also start to introduce modifications to gravity from this torsion-based formulation, in a similar fashion to the curvature-based one. Thus, in a close analogy with the $f(R)$,  the $f(T)$ gravity theory is obtained by extending the Lagrangian density of TG, that is to say, the scalar torsion $T$, to an arbitrary function of the same scalar $T$ \cite{Bengochea:2008gz,Linder:2010py}. Although  GR and TG are equivalent theories, the $f(R)$ and $f(T)$ gravity represent different modified gravity theories. In comparison with $f(R)$, whose field equations are of fourth-order, the $f(T)$ gravity has the advantage that its dynamics is given by second-order differential equations \cite{Ferraro:2006jd}. This remarkable characteristic, added to the fact that $f(T)$ gravity allow us to explain the currently observed accelerated expansion of the Universe, has given rise to a fair number of papers on these gravity theories, in which several features of $f(T)$ gravity have been examined, including observational solar system constraints~\cite{Iorio-Saridakis-2012,Iorio-2015,Farrugia-2016}, cosmological constraints~\cite{Bengochea-2011,Wei-Ma-Qi-2011,Capozziello-Luongo-Saridakis-2015,Oikonomou-Saridakis-2016,Nunes-Pan-Saridakis-2016}, dynamical behavior~\cite{Wu-Yu-b-2010}, cosmological perturbations~\cite{Chen:2010va,Dent-Duta-Saridakis-2011,Zheng-Huang-2011,Wu:2012hs,Izumi:2012qj,Li:2011wu,Cai:2015emx,Nesseris:2013jea,Basilakos:2016xob}, spherically symmetric solutions~\cite{Wang-2011,Atazadeh:2012am,Ruggiero:2015oka}, the existence of relativistic stars~\cite{Stars-in-f(T)}, cosmographic constraints~\cite{Cosmography-2011}, energy conditions bounds~\cite{Liu-Reboucas-2012}, homogeneous G\"{o}del-type solutions \cite{Liu:2012kka,Otalora:2017qqc} and gravitational waves (GWs) constraints \cite{Cai:2018rzd,Li:2018ixg}. For an excellent review on $f(T)$ gravity see Ref. \cite{Cai:2015emx} and for some others important aspects on it such as Local Lorentz invariance, see Refs. \cite{Li:2010cg,Krssak:2015oua}.

A very important generalization of $f(T)$ gravity is obtained by allowing a non-minimal coupling between torsion and matter \cite{Harko:2014aja,Farrugia:2016pjh,Harko:2014sja,Carloni:2015lsa}. This non-minimal coupling arises in a close analogy with the curvature-matter coupling in $f(R)$ gravity \cite{Nojiri:2004bi,Allemandi:2005qs,Nojiri:2006ri,Bertolami:2007gv,Harko:2008qz,Harko:2010mv,Bertolami:2009ic,Bertolami:2013kca,Wang:2013fja,Bertolami:2011fz,Bertolami:2010cw,Gomes:2016cwj,Otalora:2018bso}, whose origin can have several motivations. It is well known for example that non-minimal coupling terms acting as counterterms are required when quantizing a self-interacting scalar field in curved spacetime \cite{Birrell:1982ix}. Thus, one may be tempted to relate the need for this non-minimal coupling between gravity and matter to the existence of scalar-tensor theories and the low-energy limit of string theory \cite{Asselmeyer-Maluga:2016mvv}. In Ref. \cite{Harko:2014sja} the authors have proposed a new extension of the $f(T)$ gravity by including the coupling with an arbitrary function of the scalar tensor $T$ to the matter Lagrangian density. For the Friedmann-Robertson-Walker (FRW) background geometry, they have shown that this novel theory allows us to obtain an effective dark energy sector whose equation-of-state (EOS) parameter can be quintessence- or phantom-like, or exhibit the phantom-divide crossing, being that for a large range of the model parameters the Universe undergoes a de Sitter, dark-energy-dominated, accelerating phase. Furthermore, it can provide an early-time inflationary solution too, and hence, it is also possible an unified description of the history of cosmological expansion. On the other hand, in Ref. \cite{Carloni:2015lsa} has been studied the cosmological applications for this model from the perspective of dynamical systems by extracting the fixed points corresponding either to dark-matter-dominated, scaling decelerated solutions, or to dark-energy-dominated accelerated solutions, and thus studying their cosmological properties.

The investigation of small fluctuations around the FRW background in the framework of cosmological perturbation theory has become a cornerstone of modern cosmology. It allows us to confront any cosmological model with observations of cosmic microwave background (CMB) and large-scale structure (LSS) \cite{AmendolaTsujikawa}. Thereby, in order to reveal the full scope and predictive power of theory at hand, one must go beyond the background and enter in the perturbative level. The main goal of the present paper is to study the evolution of cosmological perturbations for this non-minimal matter-torsion coupling model \cite{Harko:2014sja,Carloni:2015lsa} in the FRW background. In particular, we are interested in the evolution of scalar perturbations and the growth of matter overdensities. 

The paper is organized as follows. In Section \ref{NMTC_Theories}, we introduce the non-minimal matter-torsion coupling theories together with the effective dark energy sector in the FRW background. In section \ref{LSPert} we define the perturbed tetrad field and obtain the corresponding linearised field equations along with the evolution equations for the matter density perturbations. In section \ref{Growth_Matter} we study the growth of matter overdensities under the quasi-static approximation for sub-horizon scales and for a power-law coupling function. Finally, in Section \ref{Concluding_Remarks}, we summarize our findings and present our main conclusions and final remarks.

\section{Non-minimal torsion-matter coupling theories}\label{NMTC_Theories}

\subsection{Field equations}

In $f(T)$ theories the dynamical variables are the tetrad fields, $\mathbf{e}^{}_{A}(x^\mu)$,
corresponding to a set of four ($A= 0,\cdots,3$) vector fields that define a local orthonormal Lorentz frame at every point $x^\mu$ of the spacetime manifold. They connect the spacetime metric (base space) $g_{\mu\nu}$ and the tangent space metric (fiber) $\eta_{AB}$ thorough the following local relation
\begin{equation}  \label{metrics}     
 g_{\mu\nu}=e^{A}_{~\mu}\,e^{B}_{~\nu}\,\eta_{AB}^{}\,,
 \end{equation} where $e^{A}_{~\mu}$ are the tetrad components in a coordinate base and they satisfy the orthogonality conditions  $e^{A}_{~\mu} e_{A}^{~\nu}=\delta^{\nu}_{\mu}$ and $e^{A}_{~\mu} e_{B}^{~\mu}=\delta^{A}_{B}$, being that the $e_{B}^{~\mu}$ are the respective inverse components. The tangent space metric, $\eta_{AB}$ and $\eta^{AB}$, lowering and raising the Lorentz indices (Latin upper-case letters) is defined as the Minkowski metric $\eta _{AB}^{}=\text{diag}\,(1,-1,-1,-1)$. On the other hand, the spacetime indices (Greek letters) vary from $0$ to $3$ and they are lowered and raised by the spacetime metric $g_{\mu\nu}$ and $g^{\mu\nu}$.

Instead of the Levi-Civita connection, one uses the Weitzenb\"{o}ck connection which is given by
\begin{equation} \label{connection}
\Gamma^{\rho}_{~\nu \mu}=e_{A}^{~\rho}\left(\partial_{\mu}{e^{A}_{~\nu}}
+\omega^{A}_{~B \mu} e^{B}_{~\nu}\right)\,,
\end{equation}
where the coefficient $\omega^{A}_{~B\mu}$ is the spin connection. This is a connection with zero curvature, but, nonzero torsion, and in components it is given by ~\cite{Aldrovandi-Pereira-book}
\begin{equation} \label{Def_Torsion}   
 T^{A}_{~\mu\nu}\equiv \partial_{\mu}e^{A}_{~\nu}
 -\partial_{\nu}e^{A}_{~\mu}+\omega^{A}_{~B\mu}\,e^{B}_{~\nu}
 -\omega^{A}_{~B\nu}\,e^{B}_{~\mu}\,.
\end{equation}

Now, if one further defines the so-called super-potential
\begin{equation} \label{Superpotential}
 S_{A}^{~\mu\nu}\equiv \frac{1}{2}\left(K^{\mu\nu}_{~~A}+e_{A}^{~\mu} \,T^{\theta\nu}_{~~\theta}-e_{A}^{~\nu}\,T^{\theta\mu}_{~~\theta}\right)\,,
\end{equation}
where
\begin{equation}  \label{Contortion}
 K^{\mu\nu}_{~~A}\equiv -\frac{1}{2}\left(T^{\mu\nu}_{~~A}
 -T^{\nu\mu}_{~~A}-T_{A}^{~\mu\nu}\right)
\end{equation}
is the contorsion tensor, we can define the torsion scalar
\begin{equation} \label{scalar}
 T \equiv S_{A}^{~\mu\nu}\,T^{A}_{~\mu\nu}\,.
 \end{equation}

In the simplest case the gravitational Lagrangian density is built by using this scalar torsion $T$ such that the relevant action is written as 
\begin{equation}
S=\int d^4x e \left[\kappa f(T)+\mathcal{L}_m\right],
\label{f(T)Gravity}
\end{equation}
where $\kappa=1/(16 \pi G)$, $e=$det$(e^{A}_{\mu})=\sqrt{-g}$, $f(T)$ is an arbitrary function of $T$, and $\mathcal{L}_m$ is the matter Lagrangian density of matter. These are the so-called $f(T)$ gravity theories whose cosmological dynamics has been studied both at background level \cite{Bengochea:2008gz,Linder:2010py,Ferraro:2006jd,Cai:2015emx}, as well at perturbations level in Ref. \cite{Wu:2012hs,Izumi:2012qj,Li:2018ixg}. The above action can also be generalized by including a non-minimal coupling between torsion and matter in the following way \cite{Harko:2014sja,Carloni:2015lsa}
\begin{equation}
S=\int
d^{4}x\,e\,\left[\kappa f_{1}(T)+f_{2}(T)\,\mathcal{L}_{m}\right],
\label{1}
\end{equation}
where $f_{i}(T)$ (with $i=1,2$) are arbitrary functions of the
torsion scalar $T$.

Varying this action with respect to the tetrad field $e^{A}_{~\mu}$ one obtains the field equation
equations
\begin{eqnarray}
&&F\Big[e^{-1} \partial_{\nu}{(e e_{A}^{~\alpha} S_{\alpha}{}^{\mu \nu})}-e_{A}^{~\alpha} T^{\lambda}{}_{\rho \alpha} S_{\lambda}{}^{\rho \mu}
+\omega^{B}_{~A\rho} e_{B}^{~\sigma} S_{\sigma}{}^{\rho\mu}\Big]\nonumber\\
&&+ \partial_{\nu}{F} e_{A}^{~\alpha} S_{\alpha}{}^{\mu\nu}+\frac{1}{4}\kappa f_{1} e_{A}^{~\mu}=\frac{1}{4} f_{2} e^{\alpha}_{A} \mathcal{T}_{\alpha}^{~\mu},
\label{geneoms}
\end{eqnarray} where we have defined $F\equiv \kappa f_{1}'+f_{2}'\mathcal{L}_{m}$ and prime denotes differentiation with respect to $T$. Here, we also have assumed that the Lagrangian matter density does not depend on derivatives of the tetrad, and the symmetric energy-momentum tensor of matter is defined as 
\be
\mathcal{T}_{\alpha}^{~\mu}=e^{A}_{~\alpha}\left[-\frac{1}{e}\frac{\delta{S}_{m}}{\delta{e}^{A}_{~\mu}}\right],
\ee with $S_{m}=\int{d^{4}x e \mathcal{L}_{m}}$ the action of matter.  Clearly, when $f_{2}(T)=1$, Eq. \eqref{geneoms} reduces to the field equations of $f(T)$ gravity \cite{Li:2018ixg}. On the other hand, GR is recovered when $f_{1}(T)=T$ and $f_{2}(T)=1$ \cite{Aldrovandi-Pereira-book}.

We can rewrite the above field equations in a covariant form by contracting them with the tetrad field $e^{A}_{~\alpha}$ as follows
\begin{eqnarray}
&&F G_{\nu}^{~\mu}+\partial_{\lambda}{F} S_{\nu}^{~\mu\lambda}+\frac{1}{4} \left[\kappa f_{1}-T F\right] \delta_{\nu}^{\mu}=\frac{1}{4}f_{2}\mathcal{T}_{\nu}^{~\mu},
\label{geneoms2}
\end{eqnarray} where
\bea
&& G_{\nu}^{~\mu}=e^{-1}  e^{A}_{~\nu} \partial_{\lambda}{(e e_{A}^{~\alpha} S_{\alpha}{}^{\mu \lambda})}- T^{\lambda}{}_{\sigma \nu} S_{\lambda}{}^{\sigma \mu}\nonumber\\
&&+e^{A}_{~\nu} \omega^{B}_{~A \sigma} e_{B}^{~\lambda} S_{\lambda}{}^{\sigma\mu}+\frac{1}{4}\delta_{\nu}^{\mu} T,
\eea is a symmetric tensor equivalent to the Einstein tensor. The right hand side of field equations \eqref{geneoms2} is symmetric, but the left hand side is not symmetric because of the local Lorentz violation in $f(T)$ gravity theories \cite{Li:2010cg}. Thus, we have an additional constraint coming from the antisymmetric part of this equation which gives
\be
\left(S_{\nu\mu}^{~~\lambda}-S_{\mu\nu}^{~~\lambda}\right)\left[F'\partial_{\lambda}T+f'_{2} \partial_{\lambda}\mathcal{L}_{m}\right]=0.
\label{Constra}
\ee Clearly, in the linear case $f_{1}(T)=T$ and $f_{2}(T)=1$ the above constraint vanishes identically, and we obtain the teleparallel equivalent of GR. Nonetheless, in the most general case, for $f(T)$ gravity and non-minimal torsion-matter coupling theories, this constraint does not vanish identically and we obtain $6$ additional equations for $6$ additional degrees of freedom \cite{Wu:2012hs}.

On the other hand, a characteristic common to non-minimal torsion (curvature)-matter coupling theories is the non-conservation of the energy-momentum tensor of matter. In a purely space-time form, by using the Bianchi identities of teleparallel gravity and the field equations \eqref{geneoms2} one obtains the non-conservation law \cite{Carloni:2015lsa}
\bea
&&\bar{\nabla}_{\mu}{\mathcal{T}_{\alpha}^{~\mu}}=-\frac{f'_{2}}{f_{2}}\left(\mathcal{T}_{\alpha}^{~\mu}+\mathcal{L}_{m}\delta^{\mu}_{\alpha}\right)\partial_{\mu}{T}+\nonumber\\
&& \frac{4}{f_{2}} K^{\rho}_{~\mu \alpha} S_{\rho}^{~\mu\nu} \partial_{\nu}{F},
\label{NonConsLAw}
\eea where $\bar{\nabla}_{\mu}$ is the covariant derivative in the Levi-Civita connection \cite{Aldrovandi-Pereira-book}. Therefore, the coupling between the matter
and torsion describes an exchange of energy and momentum
between both.

\subsection{Cosmological Background}

Proceeding forward, we impose the standard homogeneous
and isotropic geometry, that is, we consider
\begin{equation}
\label{veirbFRW}
e^A_{~\mu}={\rm
diag}(1,a,a,a),
\end{equation}
which corresponds to a flat Friedmann-Robertson-Walker
(FRW) universe with metric 
\begin{equation}
ds^2= dt^2-a^2\,\delta_{ij} dx^i dx^j \,,
\label{FRWMetric}
\end{equation}
where $a$ is the scale factor which is a function of the cosmic time $t$. In relation to the matter sector, the Lagrangian density of a perfect fluid is the energy scalar representing the energy in a local rest frame for the fluid, and therefore a possible “natural choice” for the matter Lagrangian density is $\mathcal{L}_{m}=-\rho$ \cite{Groen:2007zz}. So, it leads us to the usual expression for the energy-momentum tensor of perfect fluid
\be
\mathcal{T}_{\mu \nu}=\left(\rho+p\right) u_{\mu} u_{\nu}-p g_{\mu\nu},
\label{EMT}
\ee which is in accordance with the symmetries of the FRW spacetime defined in Eq. \eqref{FRWMetric}. From Eqs. \eqref{NonConsLAw}, it is straightforward to see that at background level the energy-momentum tensor is again conserved, just like in teleparallel gravity or $f(T)$ theories, leading to the usual continuity equation for the matter energy density
\be
\dot{\rho}+3 \gamma_{m} H \rho=0,
\label{MatterEqBack}
\ee where we have defined the parameter $\gamma_{m}\equiv 1+w_{m}$ with $w_{m}\equiv p/\rho$ the equation of state (EOS) parameter of matter. By using the tetrad field \eqref{veirbFRW} and the energy-momentum tensor \eqref{EMT} into the field equations \eqref{geneoms} we obtain the modified Friedmann equations
\bea
\label{H00}
&& 12 F H^2=f_{2} \rho-\kappa f_{1},\\
&& H \dot{F}+F \dot{H}=-\frac{1}{4} \gamma_{m} \rho f_{2}.
\label{Hii}
\eea We also have used the useful relation $T=-6 H^2$, which is obtained from the Eq. \eqref{scalar} by using the tetrad field \eqref{veirbFRW}. In the limit $F=\kappa$, that is, $f_1(T)\equiv T$, and $f_2(T)\equiv 1$, Eqs.~ \eqref{H00} and \eqref{Hii} reduce to the usual form of Friedmann equations in GR. However, the generalized Friedmann equations  \eqref{H00} and \eqref{Hii}  can be rewritten in their standard form 
\bea
\label{F00}
&& 3 H^2=\frac{1}{2\kappa} \left(\rho+\rho_{DE}\right),\\
&& -2 \dot{H}=\frac{1}{2\kappa} \left(\rho+p+\rho_{DE}+p_{DE}\right),
\label{Fii}
\eea whether one identifies the effective energy and pressure densities for dark energy as follows \cite{Carloni:2015lsa}
\bea
\label{rhoDE}
&& \rho_{DE}=-\frac{12 F H^2}{f_{2}}-\frac{\kappa f_{1}}{f_{2}}+6 \kappa H^2,\\
&& p_{DE}=\frac{4 H \dot{F}}{f_{2}}+\frac{4 F \dot{H}}{f_{2}}-4 \kappa \dot{H}-\rho_{DE}.
\label{pDE}
\eea 
Thus, one can also verify that dark energy satisfies the continuity equation 
\be
\dot{\rho}_{DE}+3 \gamma_{DE} H \rho_{DE}=0,
\label{DEConsvEq}
\ee where we have introduced, in analogy with the matter fluid, the parameter $\gamma_{DE}\equiv 1+w_{DE}$, being that $w_{DE}\equiv p_{DE}/\rho_{DE}$ is the EOS parameter of dark energy. Clearly, the equation \eqref{DEConsvEq} is consistent with the Eqs. \eqref{NonConsLAw} and \eqref{MatterEqBack}, indicating that only the total energy density $\rho_{t}=\rho_{DE}+\rho$ is conserved. 

Here, some useful cosmological parameters are the fractional densities of $\rho_{DE}$ and $\rho$ which are defined as $\Omega_{DE}\equiv \rho_{DE}/(6 \kappa H^2)$ and $\Omega_{m}\equiv \rho/(6 \kappa H^2)$. Thus, by using these parameters the Friedmann equation \eqref{H00} can be written as $\Omega_{DE}+\Omega_{m}=1$. Another important parameter is the effective EOS parameter $w_{eff}\equiv (p+p_{DE})/(\rho+\rho_{DE})$ which is related to the deceleration parameter $q\equiv -1-\dot{H}/H^2$ through the relation $q=(1/2)\left(1+3 w_{eff}\right)$, such that the accelerated expansion of the Universe occurs for $w_{eff}<-1/3$, or, equivalently, for $q<0$.

\section{Linear cosmological perturbations}\label{LSPert}

Let us consider a perturbed tetrad field whose sector of scalar perturbations is written as \cite{Wu:2012hs,Izumi:2012qj,Li:2018ixg} 
\bea
&& e^{0}_{~\mu}=\delta^{0}_{~\mu} \left(1+\psi\right)+a\delta^{i}_{~\mu} \partial_{i} {\chi},\\
&& e^{a}_{~\mu}=a \delta^{a}_{~i} \delta^{i}_{~\mu} \left(1-\varphi\right)+
\delta^{0}_{~\mu} \delta^{a}_{~i}\partial^{i}{\chi}.
\label{PertTetrad}
\eea The additional degree of freedom $\chi$ arises from the violation of local Lorentz symmetry in $f(T)$ gravity theories. This perturbed tetrad field leads to the usual line element for the scalar perturbations of the FRW spacetime in the longitudinal gauge \cite{AmendolaTsujikawa}
\be
ds^2=\left(1+2 \psi\right) dt^2-a^2 \left(1-2 \varphi\right)\delta_{ij} dx^{i} dx^{j}.
\ee

In the matter sector we are going to consider the perturbed energy-momentum tensor 
\bea
&& \mathcal{T}_{0}^{~0}=\rho+\delta \rho,\\
&& \mathcal{T}_{i}^{~0}=-\gamma_{m} \rho \delta u_{i},\\
&& \mathcal{T}_{i}^{~j}=-\left(p+c_{s}^2 \delta \rho\right)\delta^{i}_{~j}+\partial^{i}\partial_{j}{\pi},
\eea where we have introduced the sound velocity, $c_{s}^2=\frac{\delta {p}}{\delta\rho}$,  $\delta{u}_{i}$ characterizes the velocity perturbations of the fluid and $\pi$ is the so-called anisotropic stress \cite{AmendolaTsujikawa}.

Perturbing at first order the field equations \eqref{geneoms2} one obtains
\bea
&& \delta{F} G_{\nu}^{~\mu}+F \delta{G}_{\nu}^{~\mu}+\partial_{\lambda}{\delta{F}} S_{\nu}^{~\mu \lambda}+\partial_{\lambda}{F}\delta{S}_{\nu}^{~\mu \lambda}-\nonumber\\
&& \frac{1}{4}\left(f'_{2}\mathcal{L}_{m}\delta{T}+T\delta{F}\right)\delta^{\mu}_{\nu}=\frac{1}{4}\left(f'_{2}\delta{T} \mathcal{T}_{\nu}^{~\mu}+f_{2} \delta{\mathcal{T}_{\nu}^{~\mu}}\right),
\eea where $\delta{F}=F'\delta{T}+f'_{2} \delta{\mathcal{L}_{m}}$ and $\mathcal{L}_{m}=-\rho$. Thus, substituting the perturbed tetrad field \eqref{PertTetrad} one obtains the following perturbed field equations:
\bea
\label{Pert00}
&& F \left[3 H \left(H \psi+\dot{\varphi}\right)-\frac{\triangle \varphi}{a^2}\right]-3 H^2 \delta F=-\frac{1}{4} f_{2}\delta{\rho}, \\
\label{Pert0i}
&& F \left(\dot{\varphi}+H \psi\right)+\dot{F} \varphi=-\frac{1}{4} \gamma_{m} f_{2} \rho \delta{u},\\
\label{Perti0}
&& F \left(\dot{\varphi}+H \psi\right)-H \delta F=-\frac{1}{4} \gamma_{m} f_{2} \rho \delta u,\\
\label{Pertii}
&& F\left(\ddot{\varphi}+ H \dot{\psi}\right)+\left(\dot{F}+3 F H\right) \dot{\varphi}+\frac{ F \triangle(\psi-\varphi)}{3 a^2}+\nonumber\\
&& \frac{ \dot{F} \triangle \chi}{3 a}+\left(2 H \dot{F}+2 F \dot{H}+3 F H^2\right) \psi- H \delta{\dot{F}}-\nonumber\\
&& \left(\dot{H}+3 H^2\right) \delta{F}= \frac{1}{4}\left[f_{2} c_{s}^2 \delta{\rho}+\gamma_{m} \rho f'_{2} \delta{T}\right],
\eea where $\triangle=\delta^{i j}\partial_{i}\partial_{j}$, and 
\be
\delta{T}=12 H \dot{\varphi}+\frac{4 H \triangle \chi}{a}+12 H^2 \psi.
\label{deltaT}
\ee

On the other hand, the zero anisotropic stress assumption $\pi=0$ allows us to obtain
the relation
\be
\varphi = \frac{a\dot{F} \chi}{F}+\psi,
\label{AnisoStress}
\ee
and from Eq. \eqref{Constra} we find the following additional constraint 
\be
\dot{F} \varphi+H \delta F=0.
\label{Constra2}
\ee

The time and spatial components of the linear perturbed equation associated with the non-conservation law \eqref{NonConsLAw} are given by
\bea
\label{delta_rho}
&& \delta \dot{\rho}+3 H (1+c_{s}^2) \delta \rho= \gamma_{m} \rho \left(3 \dot{\varphi}-\frac{\triangle \delta \check{u}}{a} \right),\\
&& \delta{\dot{\check{u}}}+\left[(4-3 \gamma_{m}) H+\dot{\Psi}_{c}\right] \delta{\check{u}}=-\frac{1}{a}\Big(\psi+\delta{\Psi_{c}}+\nonumber\\
&& \frac{c_{s}^2 \delta \rho}{\gamma_{m} \rho}\Big),
\label{delta_u}
\eea where have introduced the definitions $\delta{u}\equiv a \delta\check{u}$, and $\Psi_{c}\equiv \log{f_{2}}$. In this last equation we also have used the constraint \eqref{Constra2}. The pair of equations \eqref{delta_rho} and \eqref{delta_u}, together with the perturbed field equations \eqref{Pert00},\eqref{Perti0}, \eqref{Pert0i}, \eqref{Pertii}, and, Eqs. \eqref{AnisoStress} and \eqref{Constra2}, constitute the complete set of perturbed cosmological equations which govern the dynamics of scalar linear perturbations in the longitudinal gauge. Henceforth, we restrict ourselves to non-relativistic matter such that $\gamma_{m}=1$ and $c_{s}^2=0$.

\section{Growth of matter density
perturbations}\label{Growth_Matter}

As usual, we will work in the Fourier space by expanding all perturbed quantities in Fourier modes, i.e. 	$X(\vec{r}, t)\sim e^{i\vec{k}\cdot \vec{r}} X(t)$, with $k=|\vec{k}|$ being the wavenumber, and such that $\triangle X=-k^2 X$. In studying the growth of matter overdensities we are interested in the sub-horizon scales with $k^2/a^2 H^2\gg 1$ \cite{AmendolaTsujikawa} . Therefore, in order to obtain the equation of matter perturbations approximately, we use the quasi-static approximation \cite{FR-reviews2}
\be
k^2 |X|/a^2 \gg H^2 |X|,\:\:\:\:\: \dot{X} \lesssim |H X|, 
\ee with $X=\psi$, $\varphi$, $\alpha$.

Under this approximation, from \eqref{deltaT} we obtain 
\be
\delta{T}\simeq -4 \frac{k^2}{a^2} \chi_{m},
\label{deltaT2}
\ee where we have defined $\chi_{m}\equiv a H \chi$.
On the other hand, from Eqs. \eqref{Constra2} and \eqref{deltaT2}, it is straightforward to see that
\be
\frac{k^2 \chi_{m}}{a^2}  \simeq -\frac{f'_{2}}{4 F'} \delta \rho.
\label{Constra3}
\ee

In this way, from the time-time equation \eqref{Pert00}, and by using Eqs. \eqref{AnisoStress}, \eqref{Constra2}, and \eqref{Constra3}, one obtains
\be
\frac{k^2 \psi}{a^2}  =\frac{f_{2}}{4 F} \left[-1+\frac{f'_{2}}{f_{2}}\frac{\dot{F}}{H F'}\right] \delta \rho.
\ee Comparing this last equation with the Poisson equation for modified gravity theories,
\be
\frac{k^2 \psi}{a^2}\simeq -4 \pi \tilde{G} \delta{\rho},
\ee one can identify the modified gravitational coupling constant $\tilde{G}$, which in this case becomes given by 
\be
\tilde{G}\simeq \frac{G f_{2}}{(F/\kappa)} \left[1-\frac{f'_{2}}{f_{2}}\frac{\dot{F}}{H F'}\right].
\label{G1}
\ee  By introducing the definition of the gauge-invariant matter
density perturbation $\delta\equiv \delta \rho/ \rho+3 H \delta{u}$, the equation for the evolution of matter overdensities, in the quasi-static approximation, is obtained from Eqs. \eqref{delta_rho} and \eqref{delta_u} in the form 
\be
\ddot{\delta}+\left(2 H +\dot{\Psi}_{c}\right)\dot{\delta}+\frac{k^2}{a^2}\left[\psi+\delta{\Psi}_{c}\right]\simeq 0.
\label{MatterEq}
\ee In the absence of coupling, the second term in the last equation, which is due only to the expansion rate of the Universe, has the effect of a frictional term slowing down the growth rate matter density perturbations \cite{AmendolaTsujikawa}. On the other hand, in the presence of non-minimal coupling, it appears an additional term $\dot{\Psi}_{c}$ which acts to reinforce (or decrease) the effect of the Hubble expansion rate \cite{Baker:2011wt}. Also, there is the extra term $\delta{\Psi}_{c}$ in the third factor, which can be interpreted as a potential, that analogously to $\psi$ and $\varphi$ in the sub-horizon approximation, it satisfies a Poisson equation 
\be
\frac{k^2}{a^2} \delta{\Psi}_{c}\simeq -4 \pi G_{c} \delta{\rho},
\label{AddPoissonEq}
\ee where $G_{c}$ becomes a coupling 'constant' \cite{Bertolami:2013kca}. Thus, by using Eqs. \eqref{deltaT2} and \eqref{Constra3} on \eqref{AddPoissonEq}, one finds that
\be
G_{c}=-\frac{4 k^2}{a^2}\frac{{f'_{2}}^2 G}{f_{2} \left(F'/\kappa\right)} .
\label{G2}
\ee

Collecting the latest results expressed in Eqs. \eqref{G1} and \eqref{G2}, the evolution equation for matter overdensities \eqref{MatterEq} takes the form
\be
\ddot{\delta}+\left(2 H +\dot{\Psi}_{c}\right)\dot{\delta}-4 \pi G_{eff}\rho \delta \simeq 0,
\label{deltaEvolEq}
\ee where $G_{eff}=\tilde{G}+G_{c}$ is established as the effective or total gravitational coupling given by
\be
G_{eff}\simeq \frac{G f_{2}}{(F/\kappa)}\left[1-\left(\frac{f'_{2}}{f_{2}}\right)\left(\frac{F'}{F}\right)^{-1}\left(\frac{\dot{F}}{H F}+\frac{4 k^2}{a^2} \frac{f'_{2}}{f_{2}}\right)\right].
\label{Geff}
\ee Clearly, for $f_{2}(T)=1$, one finds $G_{eff}=G/(F/\kappa)$, and hence, for $f_{1}(T)=T$, thus, $G_{eff}=G$. This result generalizes those previously obtained for $f(T)$ gravity theories in Refs. \cite{Zheng-Huang-2011,Wu:2012hs}, once that now the force of gravitational coupling, and therefore, the growth rate of matter overdensities, depends on both, the functions $f_{1}(T)$ and $f_{2}(T)$, and derivatives. Following Ref. \cite{Bertolami:2013kca}, the above expression for $G_{eff}$ can be rewritten as 
\be
\frac{G_{eff}}{G \Sigma}-1\simeq \frac{2 m_{2}}{m_{1}}\left[m_{1}\left(1+q\right)+\frac{3}{2} m_{2} \Omega_{m} \Sigma-\frac{2 k^2 m_{2}}{a^2 T} \right],
\label{Geff2}
\ee where it has been defined the parameters
\bea
`\label{sigma}
&& \Sigma\equiv \frac{f_{2}}{(F/\kappa)},\\
\label{m1}
&& m_{1}\equiv \frac{T F'}{F},\\
&& m_{2}\equiv \frac{T f'_{2}}{f_{2}},
\label{m2}
\eea and we also have introduced the fractional energy density of matter $\Omega_{m}=\rho/(6 \kappa H^2)$ and the deceleration parameter $q=-1-\dot{H}/H^2$. The above expression for the effective gravitational coupling $G_{eff}$ is more convenient since it shows an explicit dependence on the relevant cosmological parameters as $q$ and $\Omega_{m}$, and the new parameters $\Sigma$, $m_{1}$ and $m_{2}$, which condensate all the information about the functions $f_{1}(T)$ and $f_{2}(T)$.

\subsection{Power-law coupling function}

In order to study the evolution of matter overdensities for a specific model we are going to consider the ansatz 
\be
f_{1}(T)=T,\:\:\:\: f_{2}(T)=1+(T/T^{*})^n,
\label{Model}
\ee where $T^{*}$ is a characteristic torsion scale. 

Introducing the number of $e$-folds, $N=\log{a}$, the evolution equation \eqref{deltaEvolEq} for the matter perturbation becomes,
\be
\delta''+\left(2+\frac{H'}{H}+\Psi_{c}'\right) \delta'-\frac{3}{2} \frac{G_{eff}}{G} \Omega_{m} \delta\simeq 0,
\label{delta2}
\ee where primes denote derivatives with respect to $N$. This is the evolution equation of matter overdensities at sub-horizon scales. Below, we study the dynamics of matter overdensities governed by  Eq. \eqref{delta2} in two different physical regimes, the dark matter-dominated era and for late times in the dark energy dominance.

\subsubsection{Matter-dominated era}

During the dark-matter-dominated era, $\Omega_{DE}\ll \Omega_{m} \simeq 1$, it is natural to assume that the coupling function $f_{2}(T)=1+(T/T^{*})^n$ should only present small deviations from $1$, that is, one would consider the condition $(T/T^{*})^n \ll 1$, and for this case, it can be seen that $H'/H \simeq -3/2$ and $q\simeq 1/2$.
By using these assumptions in Eqs. \eqref{sigma}, \eqref{m1}, and \eqref{m2}, we find that 
\bea
&& \Sigma\simeq 1+\left(1-n\right) \left(\frac{T}{T^{*}}\right)^n,\\
&& m_{1}\simeq n \left(n-1\right) \left(\frac{T}{T^{*}}\right)^{n},\\
&& m_{2}\simeq n \left(\frac{T}{T^{*}}\right)^{n}.
\eea In the same way, from Eq. \eqref{Geff2}, we obtain 
\be
G_{eff}\simeq G\left[1+\frac{4 n}{1-n}\left(\frac{k^2}{a^2 T}\right) \left(\frac{T}{T^{*}}\right)^{n}\right].
\ee

Moreover, under these considerations one has that $|\Psi_{c}'|\simeq | m_{2}|\ll 1$, implying that the evolution equation \eqref{delta2} must assume the following form 
\be
f' +f^2+\frac{1}{2} f -\frac{3}{2}\left[1+\frac{4 n}{1-n}\left(\frac{k^2}{a^2 T}\right)\left(\frac{T}{T^{*}}\right)^n\right]\simeq 0,
\label{f}
\ee where we also have introduced the growth rate of matter fluctuations $f \equiv \delta'/\delta$ \cite{AmendolaTsujikawa}. Since for the matter-dominated era one usually has that $\delta \sim a$, and therefore $f\sim 1$, it is natural to assume the approximation $f'\ll f^2$. Hence, by integrating Eq. \eqref{f}, it is straightforward to obtain the growing mode
\be
f\simeq 1+\frac{12}{5}\frac{n}{1-n}\left(\frac{k^2}{a^2 T}\right)\left(\frac{T}{T^{*}}\right)^{n},
\ee or equivalently
\be
\delta\sim a^{1+\frac{12 n}{5 \left(1-n\right)\left(1-3 n\right)}\left(\frac{k^2}{a^2 T \log{a}}\right)\left(\frac{T}{T^{*}}\right)^n}.
\label{deltaDM}
\ee This result leads to the solution $\delta \sim a$ for the standard cold dark matter model when the first term in the power is dominant and the second may be neglected, that is to say, for $\left(\frac{T}{T^{*}}\right)^n\sim 0$. From Eq. \eqref{deltaDM} one can see that the factor $\frac{12 n}{5 \left(1-n\right)\left(1-3 n\right)}$ works in attenuating or enhancing the growth of matter overdensities. Although, for negative $n$ the growth rate $f$ is increased, the deviation with respect to the standard matter model is very small and it becomes smaller yet for $\left|n\right|\gtrsim 1$.

\subsubsection{Dark-energy-dominated era}

During the dark-energy-dominated era $\Omega_{m} \lesssim \Omega_{DE}$ it is  required that the coupling function is the dominant term such that $f_{2}(T)=1+(T/T^{*})^n\simeq (T/T^{*})^n$. Hence, the accelerated expansion regime of the Universe should be driven by this non-minimal coupling function. In this case, it can be shown that  $H'/H\simeq -1/p$, and $q\simeq -1+1/p$, where $p=2(1-n)/3$, and one has accelerated expansion for  $p>1$, or, equivalently, for $n<-1/2$. Thus, under these considerations and by using Eqs. \eqref{sigma}, \eqref{m1}, and \eqref{m2}, we obtain that 
\bea
&& \Sigma\simeq \frac{1}{n \Omega_{m}},\\
&& m_{1}\simeq n-1,\\
&& m_{2}\simeq n,
\eea whereas that, from Eq. \eqref{Geff}, it is also straightforward to see that
\be
\frac{G_{eff}}{G}\simeq \frac{4 n}{1-n}\left(\frac{k^2}{a^2 T}\right)\frac{1}{\Omega_{m}}.
\label{Gratio} 
\ee  In terms of the growth rate $f$, and having in account that in this case one has $\Psi_{c}'\simeq -2 n/p$, the evolution equation \eqref{delta2} takes the form
\be
f'+f^2+\left[2-\frac{3\left(1+2 n\right)}{2 \left(1-n\right)}\right] f +\frac{6 n}{n-1}\left(\frac{k^2}{a^2 T}\right)\simeq 0.
\ee For the dark-energy-dominated era we retain the second and fourth terms of this last equation, and thus the solution for the growth mode is
\be
f \simeq \sqrt{\frac{6 n}{1-n}\left(\frac{k^2}{a^2 T}\right)},
\label{fn} 
\ee and thus
\be
\delta\sim  a^{\frac{2}{\left(1+2 n\right) \log{a}}\sqrt{6 n (1-n)\left(\frac{k^2}{a^2 T}\right)}},
\label{deltaDE} 
\ee where we have the condition $n<-1/2$, which guarantees the accelerated expansion of the Universe, and $T=-6 H^2<0$. From Eq. 
\eqref{deltaDE}, it can be seen that during the dark-energy-dominated epoch with $n<-1/2$ the growth of matter perturbations is approximately frozen, such that the growth rate decays as $f \sim a^{(1+2 n)/(2 (1-n))}$. We can indeed find a relationship between the rate $f$ and the ratio $G_{eff}/G$ by using equations \eqref{Gratio} and \eqref{fn}, yielding
\be
f \simeq \sqrt{\frac{3}{2}\Omega_m \left(\frac{G_{eff}}{G}\right)},
\label{fGeff} 
\ee 
which implies that an effectively weakened gravitational coupling, i.e. $G_{eff}/G<1$ produces that matter perturbations grow slower than in $\Lambda$CDM. For a discussion of this issue in the minimally coupled case, see e.g \cite{Zheng-Huang-2011}.

Until now we have studied the asymptotic behaviour of matter density perturbations. However, for a further analysis of the transition from dark matter to dark energy dominated epochs for matter density perturbations at sub-horizon scales, numerical solving is involved.

\subsection{Numerical results}

\begin{figure}[htbp]
\begin{center}
\includegraphics[width=0.45\textwidth]{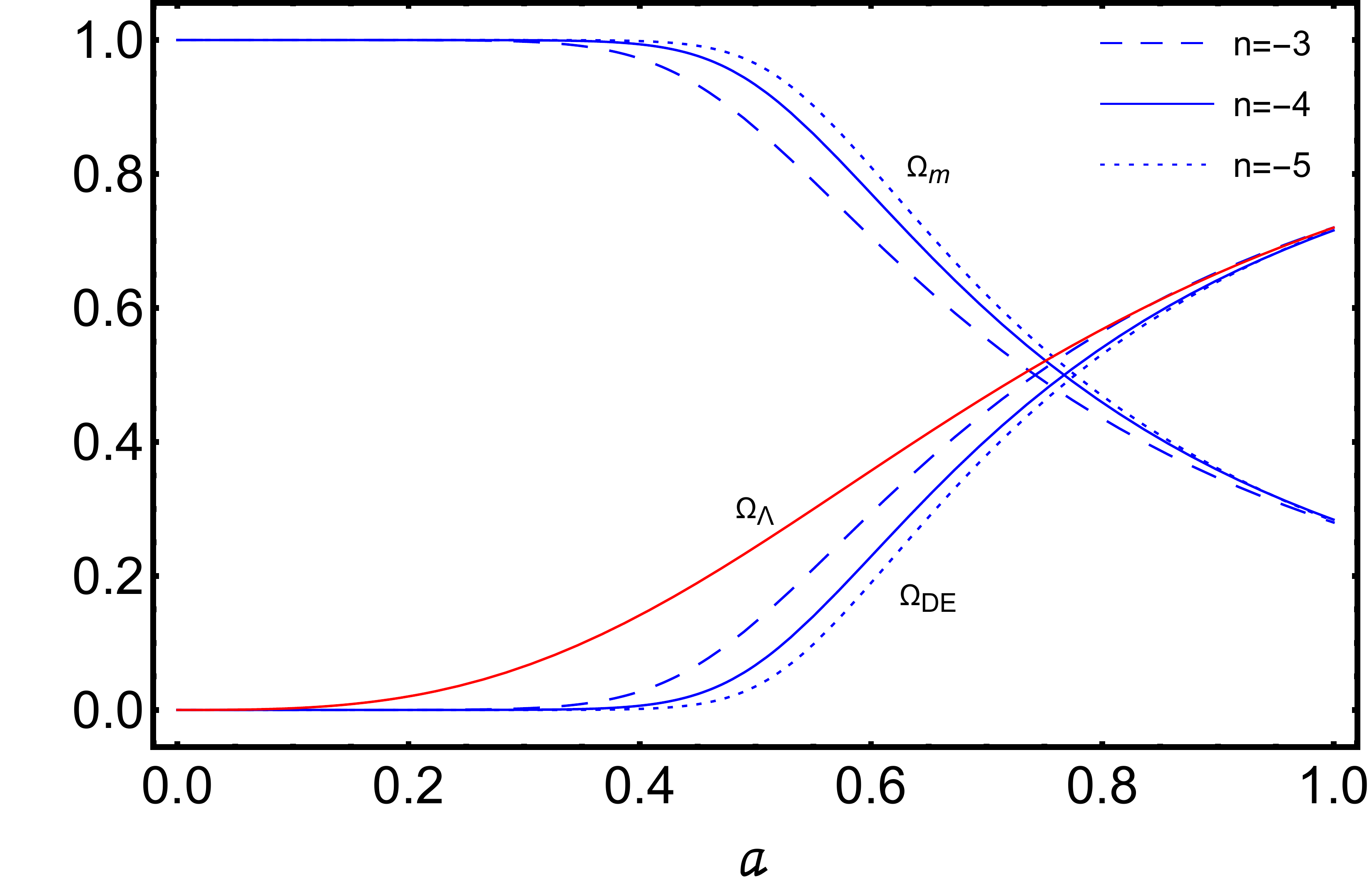}
\includegraphics[width=0.45\textwidth]{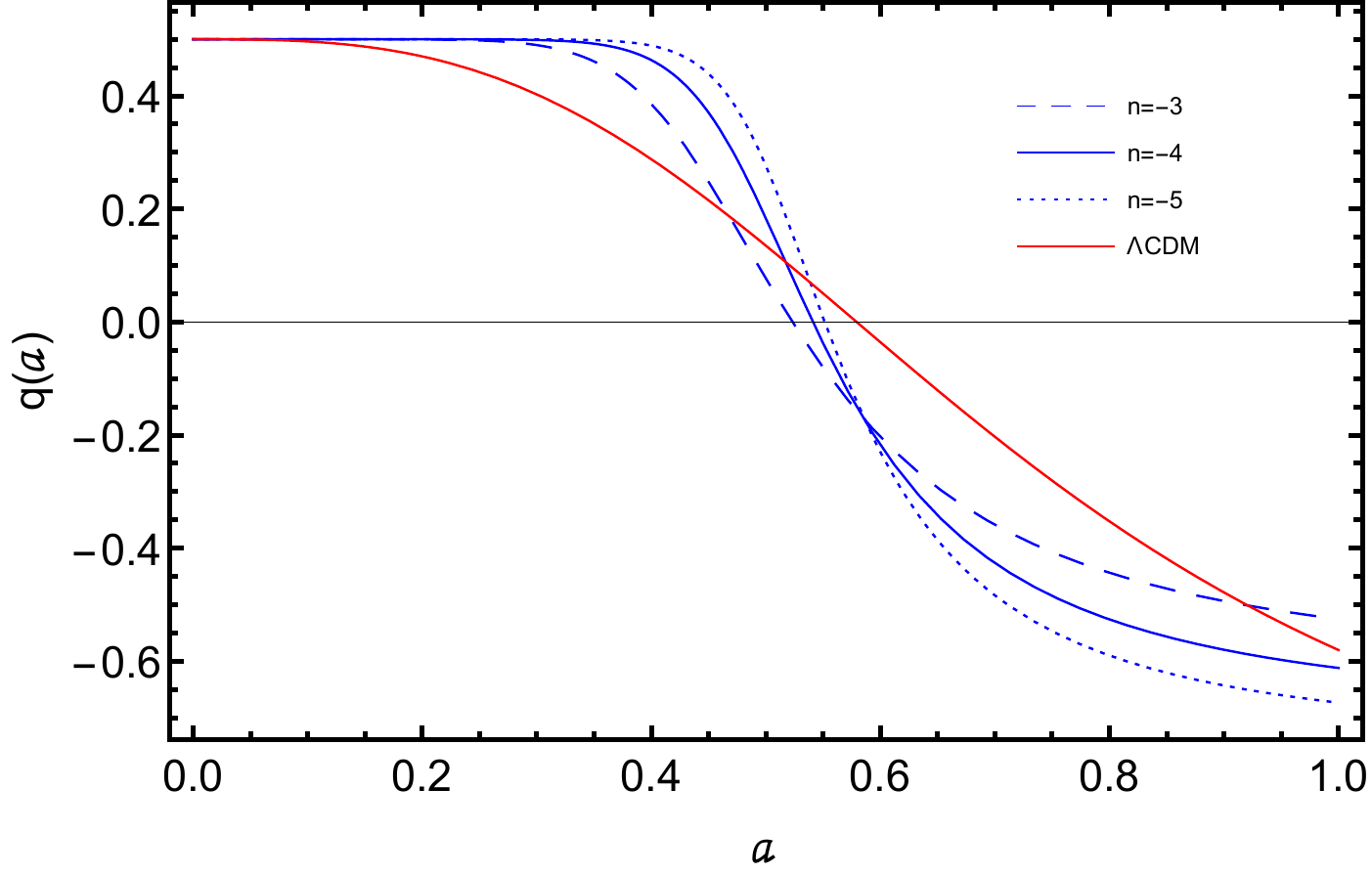}
\caption{\it{Evolution of $\Omega_{m}(a)$ and $\Omega_{DE}(a)$  (upper graph) and the deceleration parameter $q(a)$ (lower graph) as functions of the scale factor $a$ for several different values of the parameter $n$. At the present time, $a=1$ ($z=0$), the fractional densities of dark matter and dark energy take the values $\Omega_{m}(a=1)\approx 0.28$ and $\Omega_{DE}(a=1)\approx 0.72$, respectively. The Universe enters in the acceleration phase for $q(a)<0$ at $a \approx 0.56$. We also have added the $\Lambda$CDM predictions (red curve) in order to compare with the model at hand.}}
\label{FIG1}
\end{center}
\end{figure}

\begin{figure}[htbp]
\begin{center}
\includegraphics[width=0.45\textwidth]{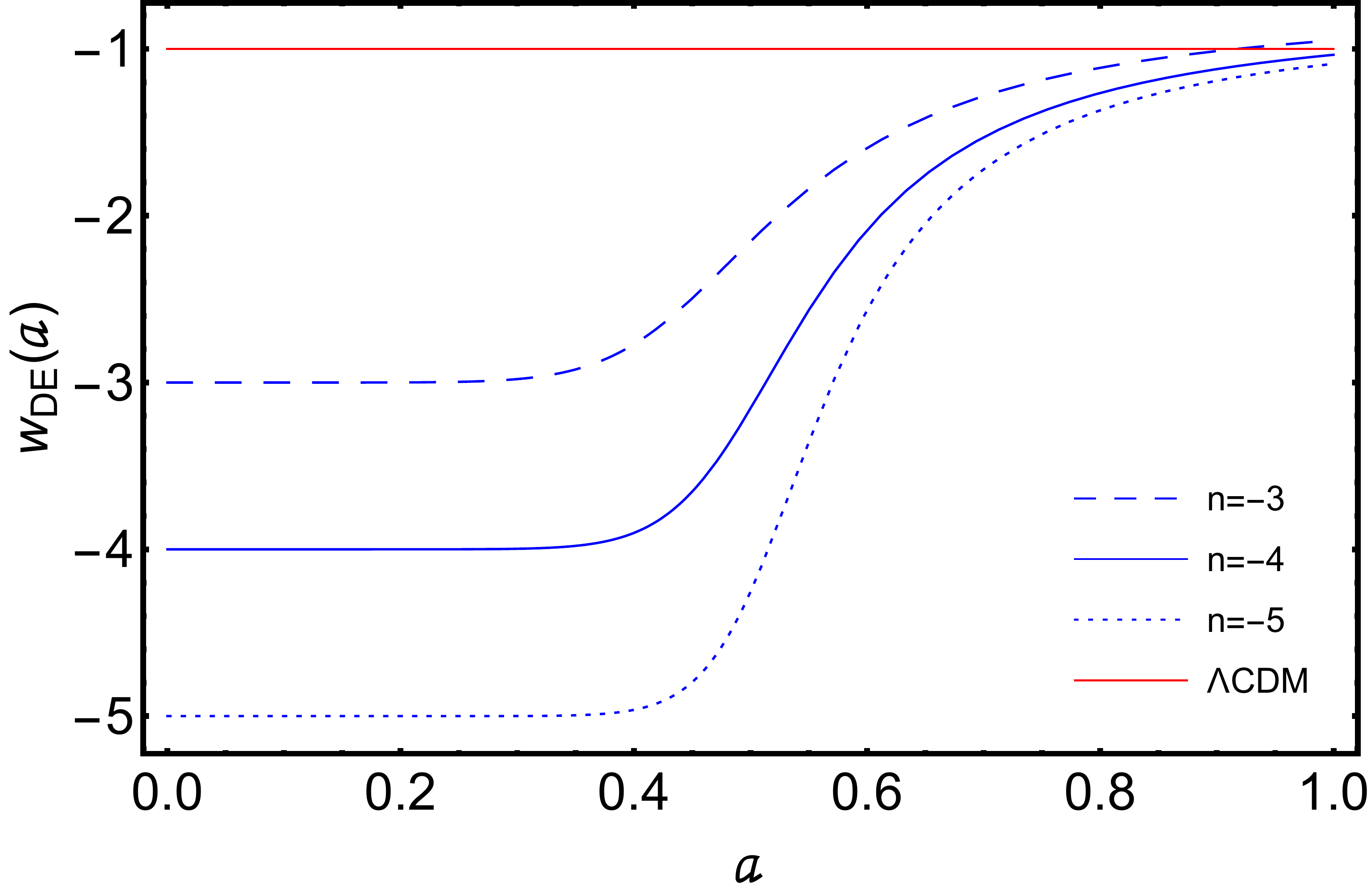}
\caption{\it{Evolution of the dark energy EOS parameter $w_{DE}(a)$ as a function of the scale factor. It is observed a phantom behaviour along of the cosmic evolution, being that the deviation for higher negative values occurs in the subdominant phase of dark energy. At the present epoch, when dark energy becomes the dominant component, $w_{DE}(a)$ may take values which are consistent with observations. For $n=-4$ one obtains $w_{DE}(a=1)\approx-1.04$ which is in accordance with observational data, whereas that for $n=-3$ and $n=-5$ one has that $w_{DE}(a=1)\approx-0.949$ and $w_{DE}(a=1)\approx-1.09$, and these values are outside of the observational bound. The red curve corresponds to the $\Lambda$CDM model.}}
\label{FIG2}
\end{center}
\end{figure}

In order to solve numerically the complete evolution of matter density perturbations for sub-horizon scales one must solve simultaneity both the cosmological background equations \eqref{F00}, \eqref{Fii}, and the matter perturbation equation \eqref{delta2}. Thus, let us introduce the new dimensionless variable $y=H/H^{*}$. In terms of this new variable and for the model \eqref{Model}, the cosmological parameters $\Omega_{m}$, $q$, and $w_{DE}$ can be written as 
\bea
\label{Omegafy}
&& \Omega_{m}(y(a))=\frac{1}{(1-2 n) y(a)^{2 n}+1},\\
\label{qfy}
&& q(y(a))=-\frac{3}{2 \left[\frac{n}{(2 n-1) y(a)^{2 n}-1}+n-1\right]}-1,\\
&& w_{DE}(y(a))=\frac{n\left[1-(2 n-1) y(a)^{2 n}\right]}{(n-1) (2 n-1) y(a)^{2 n}+1}.
\label{wDEfy}
\eea
For the ansatz \eqref{Model} one can easily see that the equations \eqref{F00}, \eqref{Fii} lead to the relation
\be
\frac{a}{y}\frac{dy}{da}=\frac{3}{2 \left[\frac{n}{(2 n-1) y^{2 n}-1}+n-1\right]}.
\label{yEq}
\ee Hence, by solving numerically Eq. \eqref{yEq} we obtain the evolution of the cosmological parameters in Eqs. \eqref{Omegafy}, \eqref{qfy} and \eqref{wDEfy}. For consistency with current observational data it is required at the present time $z=0$ that $\Omega_{m}^{(0)}\approx 0.28$, $\Omega_{DE}^{(0)}\approx 0.72$, and $w_{DE}^{(0)}=-1.028 \pm 0.032$ \cite{Aghanim:2018eyx}. In FIG \ref{FIG1} we have depicted the cosmological behaviour of $\Omega_{m}(a)$, that satisfy the constraint $\Omega_{m}(a)+\Omega_{DE}(a)=1$, and the deceleration parameter $q(a)$, as functions of the scalar factor $a$. On the other hand, in FIG \ref{FIG2} we have depicted the evolution of $w_{DE}(a)$. For several values of the power $n$ we have shown the behaviour of these cosmological parameters. For $n=-3$ and $n=-5$, one has that $w_{DE}(a=1)\approx-0.949$ and $w_{DE}(a=1)\approx-1.09$, and these values are outside of the observational bound, whereas that for $n=-4$, we find $w_{DE}(a=1)\approx-1.04$, which is consistent with current observations.

The effective EOS parameter $w_{eff}$ is related to the fractional dark energy density $\Omega_{DE}$ through the equation $w_{eff}=w_{m}+\Omega_{DE} (w_{DE}-w_{m})$. Once that $\Omega_{DE}$ falls rapidly with increasing redshift, the effective parameter $w_{eff}$ tends to the asymptotic value $w_{m}$. In FIG \ref{FIG1} (upper graph) we also depict the behaviour of $\Omega_{DE}(a)$ for the model \eqref{Model}, which becomes very small at $a\leq 0.5$ ($z\geq 1$). Despite the dark energy EOS parameter $w_{DE}$ takes large negative values at $a\leq 0.5$ (FIG \ref{FIG2}), it is still guaranteed that $\Omega_{DE} w_{DE} \ll \Omega_{m} w_{m} $ and therefore $w_{eff}\approx w_{m}$, which is compatible with observations \cite{Weinberg:2012es}.

The transition to acceleration phase is shown in FIG \ref{FIG1} (lower graph), as described by the deceleration parameter $q(a)$. As it can be observed, for model \eqref{Model} this transition occurs at around of $a\approx 0.56$ ($z\approx 0.8$). This result is in agreement with the predicted value by $\Lambda$CDM model and that one obtained in Ref. \cite{Capozziello-Luongo-Saridakis-2015} for the $f(T)$ cosmology.

The evolution equation of matter density perturbations \eqref{delta2} takes the following form

\bea
a^2 \frac{d^2\delta}{da^2}=-a\left[3+\frac{a}{y}\left(1+2 m_{2}\right)\frac{dy}{da}\right] \frac{d\delta}{da}+\frac{3}{2} \frac{G_{eff}}{G} \Omega_{m} \delta, 
\label{delta3}
\eea where $\Omega_{m}(y(a))$ is given by \eqref{Omegafy}. From Eqs. \eqref{sigma}, \eqref{m1}, \eqref{m2} one obtains
\bea
&& \Sigma(y(a))=\frac{y(a)^{2 n}+1}{n y(a)^{2 n} \Omega_{m}(a)+1},\\
&& m_{1}(y(a))=\frac{n (n-1) y(a)^{2 n} \Omega_{m}}{n y(a)^{2 n} \Omega_{m}(a)+1},\\
&& m_{2}(y(a))=\frac{n y(a)^{2 n}}{y(a)^{2 n}+1}, 
\eea and therefore $G_{eff}(y(a))$ is given by \eqref{Geff2}.

\begin{figure}[htbp]
\begin{center}
\includegraphics[width=0.45\textwidth]{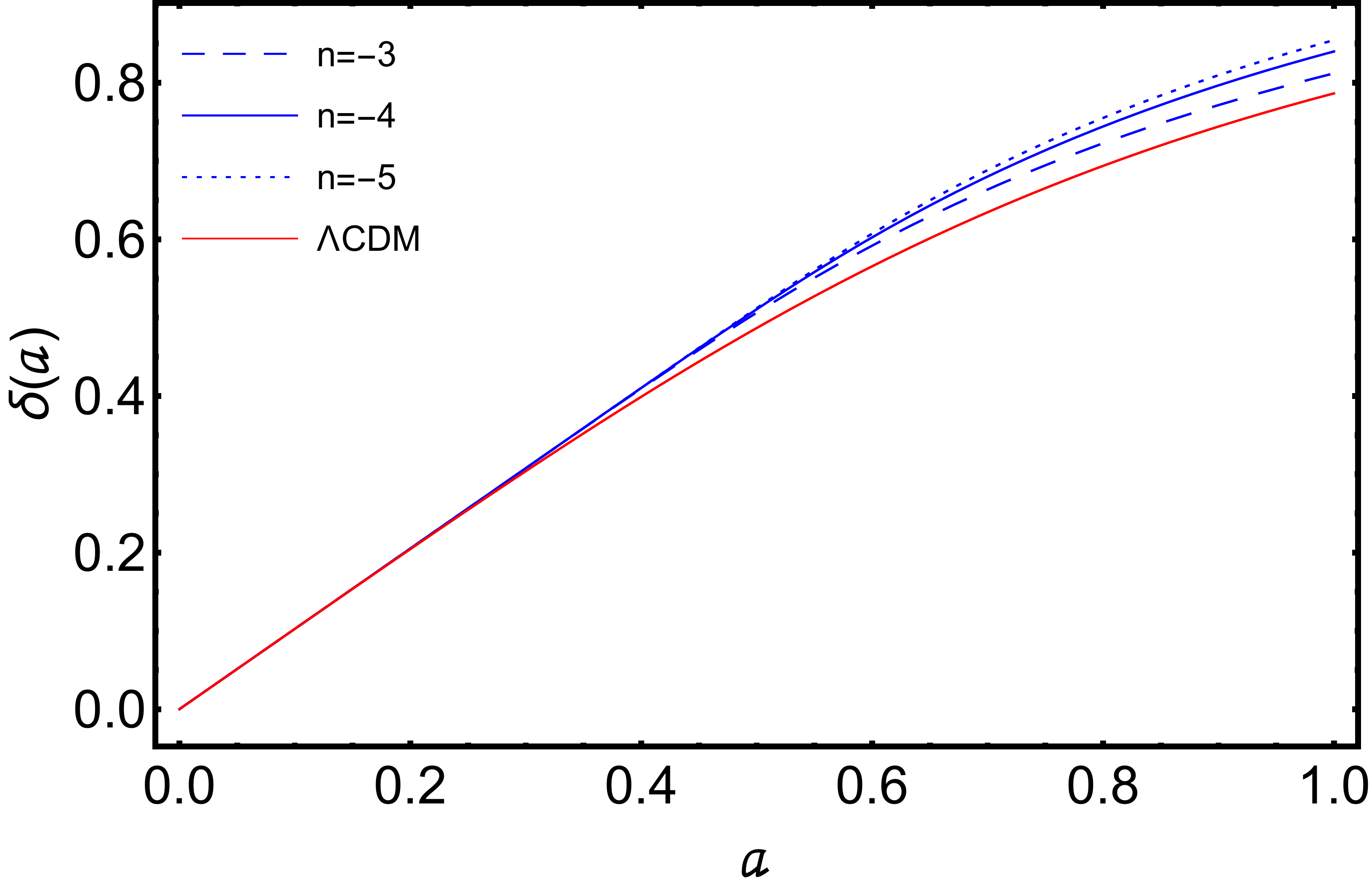}
\includegraphics[width=0.45\textwidth]{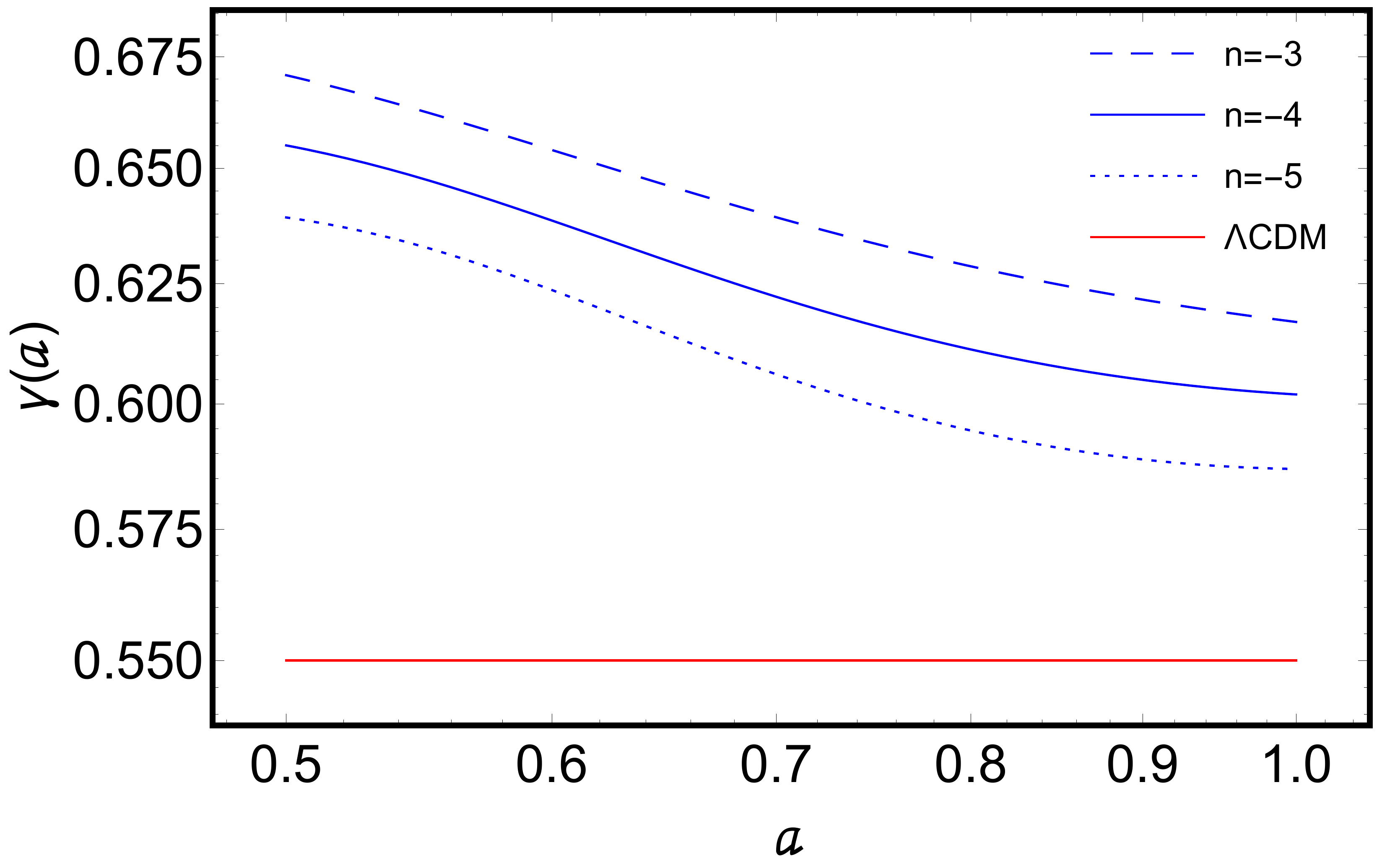}
\caption{\it{In the upper graph we shown the evolution of matter density perturbation $\delta(a)$ as a function of scale factor and for several different values of the parameter $n$, at $k=0.1 h$ $Mpc^{-1}$. In lower graph it is depicted the corresponding behaviour of the growth index $\gamma(a)$. We find a deviation with respect to the standard value $\gamma=0.55$ for the $\Lambda$CDM, being that our $\gamma(a)$ assumes values larger than it, and therefore implying a smaller growth rate.}}
\label{FIG3}
\end{center}
\end{figure}

By simultaneously integrating the differential equations \eqref{yEq} and \eqref{delta3}, we obtain the evolution for the matter density perturbation $\delta(a)$. Also, with the aim of comparison 
with $\Lambda$CDM model, we introduce the growth index $\gamma$ defined by the relation $f=\Omega_{m}(a)^{\gamma}$ \cite{Wang:1998gt}. In the case of $\Lambda$CDM model the growth index has been found to be $\gamma_{\Lambda \textup{CDM}}=6/11\simeq 0.55$ \cite{Linder:2005in,Nesseris:2007pa,Belloso:2011ms,Amendola:2016saw}. In FIG \ref{FIG3} we have depicted the numerical solution for the matter overdensity $\delta$ (upper graph) and for the growth index $\gamma$ (lower graph), computed by $\gamma=\ln f/\ln \Omega_m$, for several different values of power $n$ at 
a scale $k=0.1 h$ $Mpc^{-1}$ of the linear regime. Being consistent with the early standard matter era, from the upper graph, one observes that for the three values of the power $n$, the matter overdensity $\delta(a)$ grows linearly with the scale factor $a$ during the dark matter dominated epoch. Once that dark energy comes to dominate the dynamics at late times, the growth of matter overdensity starts to slow down with the scale factor as the power $n$ increases. On the other hand, as it can be seen from lower graph of FIG \ref{FIG3}, $\gamma$ is found to be a decreasing function with the scale factor and as the power $n$ decreases, the curve approaches the curve corresponding to $\Lambda$CDM (solid black line). Nevertheless, for values of the power $n$ such that $n<-4$, the EOS parameter for dark energy $w_{DE}$ lies outside the current observational bound, then our model predicts a growth index larger than those obtained for $\Lambda$CDM model, indicating therefore a smaller growth rate. Such a deviations from $\Lambda$CDM have been already observed in a broad class of viable $f(T)$ models in the minimally coupled case \cite{Chen:2010va,Dent-Duta-Saridakis-2011,Zheng-Huang-2011,Wu:2012hs,Izumi:2012qj,Li:2011wu,Cai:2015emx,Nesseris:2013jea,Basilakos:2016xob}. Hence observationally determining $\gamma$ allows us to distinguish between several $f(T)$ models.

In direct comparison of this non-minimal extension of $f(T)$ with its counterpart in $f(R)$ gravity, in \cite{Bertolami:2013kca,Wang:2013fja} the authors studied the effects on the evolution of matter perturbations of a curvature-matter coupling. In particular, they considered a power-law coupling function, i.e. $f_2(R)=1+(R/R_*)^n$, with $R_*$ being a characteristic curvature scale. It was found that, in consistency with current observational constraints, negative values for the power $n$ are favoured, as $n=-4$ and $n=-10$, accordingly with \cite{Bertolami:2010cw}. However, in the present work we have found that a consistency with current observations requires that the power $n$ lies in a small range about  $n=-4$. It what concerns the value $n=-10$, it yields a growth index $\gamma$  
very close to those in $\Lambda$CDM, but a present value for the EOS of dark energy being outside the allowed current bound. In this sense, a non-minimal extension of $f(T)$ gravity becomes more constrained than those in $f(R)$ gravity.

\subsection{Theoretical predictions of $f\sigma_{8}(z)$}

In order to compare further the predictions of our model with observations we are going to introduce the observable quantity $f\sigma_{8}(z)$ which is defined via the relation
\be
f\sigma_{8}(z)\equiv f(a)\cdot \sigma(a)=\frac{\sigma_{8}}{\delta{(1)}} a \delta'(a), 
\ee being that $f(a)=d\ln\delta(a)/d\log(a)$ is the growth rate and $\sigma(a)=\sigma_{8} \delta(a)/\delta(1)$. The sigma function $\sigma(a)$ is the temporal evolution of the root mean square mass fluctuation amplitude in spheres of size $8~h^{-1}~Mpc$ ($k\sim k_{\sigma_{8}}=0.125$~$h~Mpc^{-1}$) with $\sigma_{8}=\sigma(1)$, whose value is strongly depending on the physics of the late-time expansion and therefore on the specific dark energy model \cite{Sola:2017znb}. 

Currently exist some significant and persist tensions between data sets in the context of $\Lambda$CDM, which involve relevant parameters such as the Hubble parameter $H_{0}$ and $\sigma_{8}$ \cite{Aghanim:2018eyx,Sola:2017znb,Kazantzidis:2018rnb,Albarran:2016mdu}. The so-called $\sigma_{8}$-tension has its origin in the fact that the values $f\sigma_{8}$ from LSS structure formation data lie some $\sim 8$ per cent below the $\Lambda$CDM prediction. So, this seems to indicate us that the predicted value for $\sigma_{8}$ from $\Lambda$CDM it is much bigger than it should be, for the same growth rate in its present day value $f(1)$ \cite{Gomez-Valent:2018nib}. Nevertheless, it is clear that the growth rate $f(a)$ is also very dependent on the model \cite{FR-reviews2} and a lower growth rate also could generate a better fit with LSS data.

An expression for the $\sigma(a)$ function which captures all the physical information encoded in it is given by $\sigma(a)^2=\delta(a)^2 \int{d^{3}k/(2 \pi)^3 P(k,\textbf{p}) W^2(k,R_{8})}$, where $W$ is a top-hat smoothing function, $P(k,\textbf{{p}})=P_{0} k^{n_{s}} T^2(k,\textbf{p})$ is the linear matter power spectrum, $P_{0}$ is its normalization factor, and $T(k,\textbf{p})$ is the matter transfer function, with $\textbf{p}$ being the vector that contains the parameters of the model \cite{Sola:2017znb,Gomez-Valent:2018nib}. For our purposes and in view of the fact that $\sigma(a)\sim \delta(a)$ \cite{Kazantzidis:2018rnb}, from FIG \ref{FIG3} we estimate for model \eqref{Model}, with $n=-3,-4,-5$, the corresponding values $\sigma_{8}\simeq 0.81,0.84,0.85$, whereas that for $\Lambda$CDM model one obtains $\sigma_{8}\simeq 0.79$, being that we have kept the fixed values $\Omega^{(0)}_{m}\simeq 0.28$ and  $\Omega^{(0)}_{DE}\simeq 0.72$. These values become a good approximation in agreement with the values presented in Refs. \cite{Aghanim:2018eyx,Sola:2017znb,Kazantzidis:2018rnb,Albarran:2016mdu,Gomez-Valent:2018nib,Abedi:2018lkr}.

\begin{figure}[htbp]
\begin{center}
\includegraphics[width=0.45\textwidth]{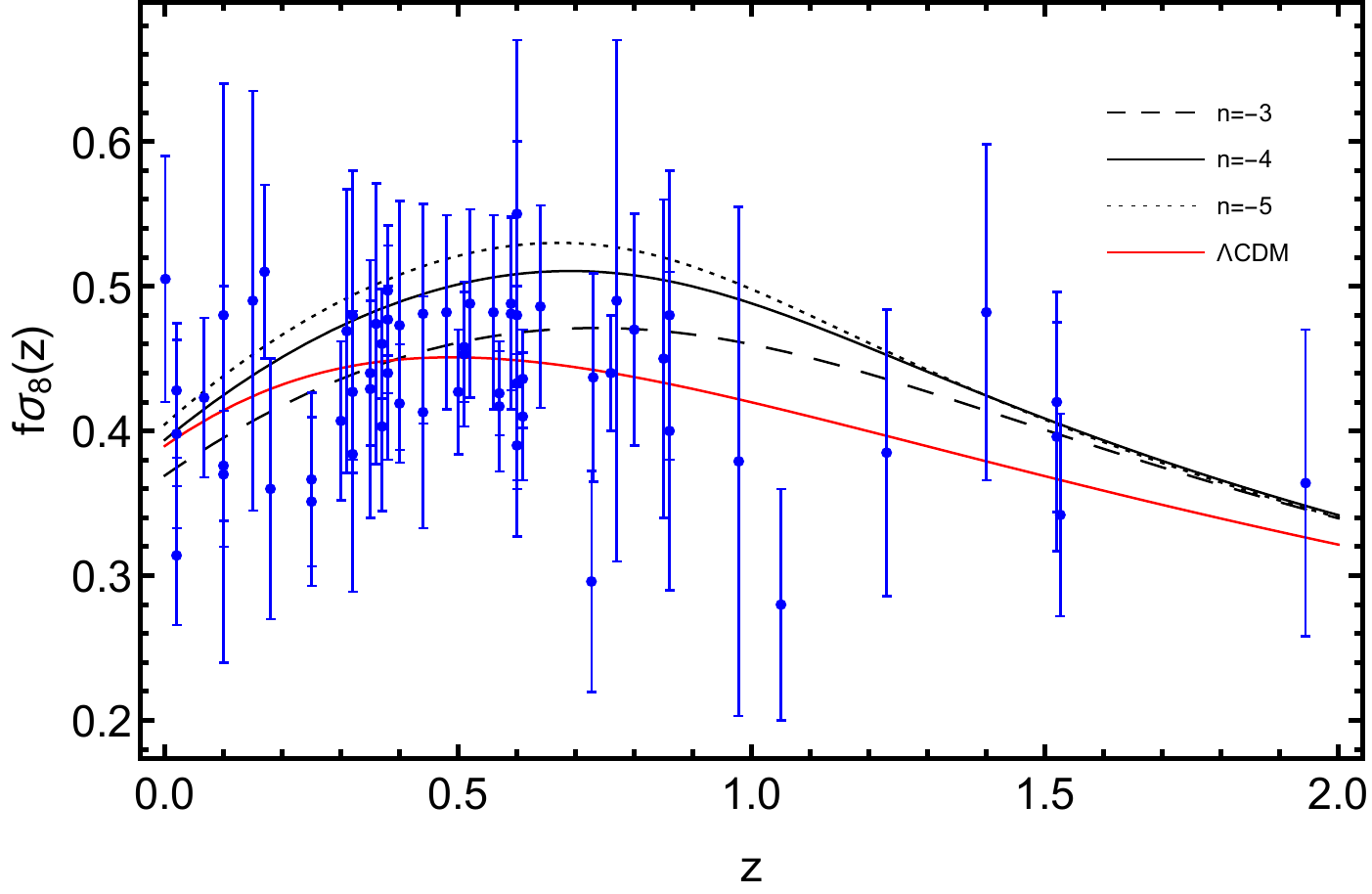}
\caption{\it{It is shown the evolution of the weighted growth rate $f\sigma_{8}(a)$ for several different values of the parameter $n$ and for $\Lambda$CDM model. In the plot we have used the full RSD $f\sigma_{8}$ dataset of Table II in Ref. \cite{Kazantzidis:2018rnb}. The prediction for $n= -3$ lies some $\sim 5$ per cent below the $\Lambda$CDM prediction, and therefore alleviating the $\sigma_{8}$-tension.}}
\label{FIG4}
\end{center}
\end{figure}

In FIG \ref{FIG4} we depict the theoretical curves for the weighted linear growth rate $f\sigma_{8}(z)$ for each model. It can be observed that the corresponding prediction for the $f\sigma_{8}(z)$ function in the case of $n=-3$ is below the respective prediction of the $\Lambda$CDM scenario and therefore the model at hand is potentially capable in alleviating the $\sigma_{8}$-tension. In order to provide a measure of this result, we introduce the exact relative difference $\Delta_{f\sigma_{8}}(z)\equiv 100\left(f\sigma_{8}(z)\mid_{model}-f\sigma_{8}(z)\mid_{\Lambda CDM}\right)/f\sigma_{8}(z)\mid_{\Lambda CDM}$ with respect to the concordance model \cite{Gomez-Valent:2018nib}. In FIG \ref{FIG4}, for $n=-3$ one obtains $f\sigma_{8}(0)\approx 0.37$, whereas that for $\Lambda$CDM it is seen that $f\sigma_{8}(0)\mid_{\Lambda CDM} \approx 0.39$, which gives the relative difference $\Delta_{f\sigma_{8}}(0)\approx 5$. Hence, the prediction for $n= -3$ lies some $\sim 5$ per cent below the $\Lambda$CDM prediction. Also, it is worth noting that to accommodate the observational limits of the dark energy EOS parameter at the present time $z=0$, one could take a value of $n$ slightly smaller than $-3$, and one could still have a relative (percentage) difference in approximately $\sim 4-5$ per cent with respect to $\Lambda$CDM. It can also be highlighted that this result improves that one obtained for the so-called XCDM parametrization in Ref. \cite{Gomez-Valent:2018nib}.

\section{Concluding remarks}\label{Concluding_Remarks}
In studying the dark energy problem of cosmology a very interesting class of modified gravity models are the so-called $f(T)$ gravity theories \cite{Bengochea:2008gz,Linder:2010py,Ferraro:2006jd}, that generalize the teleparallel equivalent of GR \cite{Einstein,TranslationEinstein,Early-papers1,Early-papers2,Early-papers3,Early-papers4,Early-papers5,Early-papers6}, in which gravity is described through torsion and not curvature \cite{Aldrovandi-Pereira-book,JGPereira2,AndradeGuillenPereira-00,Arcos:2005ec}. These torsion-based modified gravity theories constitute a good alternative to the conventional based-curvature modified gravity models \cite{Cai:2015emx}. Furthermore, in the same spirit of non-minimal $f(R)$ gravity theories \cite{Nojiri:2004bi,Allemandi:2005qs,Nojiri:2006ri,Bertolami:2007gv,Harko:2008qz,Harko:2010mv,Bertolami:2009ic,Bertolami:2013kca,Wang:2013fja,Bertolami:2011fz,Bertolami:2010cw,Gomes:2016cwj,Otalora:2018bso}, one may think in an attractive generalization of this framework, by allowing a non-minimal coupling between matter and torsion \cite{Harko:2014sja}. This generalized theory has proven to have very important features at background level, providing an explanation for both late-time accelerated expansion, and early-time inflationary phase, that in this way, it also could lead to a possible unified description of the cosmological expansion history \cite{Carloni:2015lsa}. Additionally, it is fundamental a study of cosmological perturbations in order to compare all the predictions and results obtained from the model with the observational data of cosmic microwave background (CMB) and large-scale structure (LSS) \cite{AmendolaTsujikawa}. 

In the present paper we have studied the evolution of scalar cosmological perturbations in these non-minimal torsion-matter coupling theories. In particular, by using the quasi-static approximation in sub-horizon scales we have obtained the evolution equation for matter density perturbations as written in Eq. \eqref{MatterEq}. Thus, we have found an effective gravitational coupling ’constant’ $G_{eff}$ given by Eq. \eqref{Geff}, which carries an additional contribution $G_{c}$, as defined in Eq. \eqref{G2}, whose origin is related to the non-minimal matter-torsion coupling function. This result constitutes a generalization of those previously obtained for the growth of matter overdensities in minimal $f(T)$ gravity, since the strength of the gravitational coupling, as given by $G_{eff}/G$, now depends on both functions, $f_{1}(T)$ and $f_{2}(T)$, and their derivatives. In Eq. \eqref{Geff2} we have rewritten  $G_{eff}$ in terms of the relevant cosmological parameters, the deceleration parameter $q$ and the fractional matter density $\Omega_{m}$, and the new parameters $\Sigma(T)=f_{2}/(F/\kappa)$, $m_{1}(T)=T F'/F$ and $m_{2}(T)=T f'_{2}/f_{2}$, which encode all the information about the model in the functions $f_{1}(T)$ and $f_{2}(T)$. Clearly, in order to decide something about the behaviour of the ratio $G_{eff}/G$, it is necessary to know the specific functional form of this set of parameters.

By applying our results to a particular model, we have considered the important case of a power-law coupling function $f_{2}(T)=1+(T/T^{*})^{n}$, with $n$ negative and $T^{*}$ a characteristic torsion scale, which we have assumed to be the dominant term in late-times, and hence leading the accelerated expansion of the Universe in the dark energy regime. Here, to isolate the effects of the non-minimal coupling between torsion and matter we have taken the pure gravitational sector to have the teleparallel equivalent form of GR, that is to say, $f_{1}(T)=T$.
With these assumptions we have analytically solved the evolution equation of matter overdensities in Eq \eqref{delta2}, in two different asymptotic regimes through cosmological evolution, the dark matter-dominated era, $(T/T^{*})^{n}\ll 1$, and dark energy-dominated epoch, $(T/T^{*})^{n}\gg 1$. We have obtained the analytic solution for the matter overdensity \eqref{deltaDM} validates for the dark matter-dominated regime, which allows us to compare with the solution $\delta\sim a$ for the standard cold dark matter era. Clearly, our solution \eqref{deltaDM} reproduces this linear growth with the scalar factor in the limit  $(T/T^{*})^{n}\rightarrow 0$, but more importantly, one can see that the factor $12 n/\left[5 \left(1-n\right)\left(1-3 n\right)\right]$ works in attenuating or enhancing the growth of matter overdensities. However, although for negative $n$ the growth rate $f=\delta'/\delta$ is increased, the deviation with respect to the standard matter model is very small and it becomes smaller yet for $\left|n\right|\gtrsim 1$. On the other hand, for the dark energy regime we have found the analytic solution \eqref{deltaDE}. This solution allows us to show that for $n<-1/2$ the growth of matter perturbations is approximately frozen, such that the growth rate decays as $f=\delta'/\delta \sim a^{(1+2 n)/(2 (1-n))}$. An interesting additional conclusion about this solution can be obtained if one puts the growth rate $f$ in terms of the ratio $G_{eff}/G$ as in Eq. \eqref{fGeff}. This relation implies that an effectively weakened gravitational coupling, i.e. $G_{eff}/G<1$, produces a growth of matter overdensities slower than in $\Lambda$CDM, as it has been further corroborated by our results in a subsequent  numerical analysis. 

Since the transition between the dark matter and dark energy-dominated eras is not included in the above analytical analysis, a complete numerical analysis is also required. Thus, we have numerically solved the matter perturbation equation \eqref{delta3}. In this way, from the numerical solution for the matter perturbation $\delta$ (FIG \ref{FIG3} (upper graph)) we have computed the growth index $\gamma=\ln f/\ln \Omega_m$ (FIG \ref{FIG3} (lower graph)), for several different values of power $n$ at a scale $k=0.1 h$ $Mpc^{-1}$ of the linear regime. From this numerical analysis we have ratified all the results previously obtained in the analytical analysis. Also, we have found that our model predicts a growth index larger than those obtained for $\Lambda$CDM model, indicating therefore a smaller growth rate. In full agreement with these results, we also have found that the model at hand is potentially capable in relaxing the existing $\sigma_{8}$-tension, once that it can provide us a $f\sigma_{8}$ prediction which is $\sim 4-5$ per cent below the corresponding $\Lambda$CDM prediction.
Furthermore, after comparing with its counterpart in $f(R)$ gravity, where a non-minimally coupling of power-law form between matter and curvature has been studied \cite{Bertolami:2013kca,Wang:2013fja}, we found that a non-minimal power-law coupling between matter and torsion in $f(T)$ gravity becomes more constrained than those in $f(R)$ gravity. 

The explicit coupling between the torsion scalar and the matter Lagrangian density has as consequence an energy exchange between matter and gravity, which manifests itself in the non-vanishing of the covariant divergence of the matter stress-energy tensor as it is shown in Eq. \eqref{NonConsLAw}. This non-conservation of energy can be interpreted as a failure of the theory in relation to the so-called metric postulates \cite{C_M_Will}, as it could generate a non-geodesic motion of test bodies, and therefore it also could imply a possible violation of Einstein equivalence principle (EEP) \cite{Bertolami:2007gv}. Furthermore, in Ref. \cite{Bertolami:2007gv} it was also suggested that for the parametrization $f_{2}(R)=1+\lambda \tilde{f}_{2}(R)$ of the non-minimal coupling function between curvature and matter one could potentially tune the parameter $\lambda$ with the purpose of reducing the effects of such violation below current experimental accuracy. Nevertheless, as it has also been shown in Ref. \cite{Sotiriou:2008it}, the metric postulates or the non-conservation of energy do not themselves provide quantitative estimates of the deviations from the EEP. So, in order to decide something with respect to the relationship between the values of the parameters $n$, and $T^{*}$, for the parametrization $f_{2}(T)=1+\left(T/T^{*}\right)^n$ of non-minimal torsion-matter coupling function, and the measured bounds of the EEP, a more detailed study must be performed in this direction. This necessary study lie beyond the scope of the present work, and is left for a separate project.

Finally, it is important to highlight that due to the Local Lorentz violation in $f(T)$ gravity, and its extensions, one has an extra degree of freedom represented by the scalar perturbation $\chi$ in the perturbative framework developed for these theories \cite{Zheng-Huang-2011,Wu:2012hs,Izumi:2012qj,Li:2018ixg}. As it also happens in the miniminal case, we have shown that in the non-minimal extension of $f(T)$ gravity, this additional scalar perturbation $\chi$ does not have a significant contribution on the growth of matter overdensities at sub-horizon scales. However, as it has been shown in Refs. \cite{Wu:2012hs,Li:2011wu} for the $f(T)$ gravity theory, it is expected that at super-horizon scales this new scalar mode has an important effect on the evolution of matter perturbation.

\begin{acknowledgements}
G. Otalora acknowldeges DI-VRIEA for financial support through Proyecto Postdoctorado $2018$ VRIEA-PUCV. The author N.V. was supported by Comisi\'on Nacional de Ciencias y Tecnolog\'ia of Chile through FONDECYT Grant N$^{\textup{o}}$ 11170162.
M. Gonzalez-Espinoza acknowledges support from a PUCV doctoral scholarship.
\end{acknowledgements}



\begin{thebibliography}{99}

\bibitem{Riess:1998cb} 
A.G. Riess {\it et al.} [Supernova Search Team], Observational evidence from supernovae for an accelerating universe and a cosmological constant. Astron.\ J.\  {\bf 116}, 1009 (1998)
 
\bibitem{Perlmutter:1998np} 
S. Perlmutter {\it et al.} [Supernova Cosmology Project Collaboration], Measurements of Omega and Lambda from $42$ high redshift supernovae. Astrophys.\ J.\  {\bf 517}, 565 (1999)
  
\bibitem{Ade:2013zuv} 
P.A.R. Ade {\it et al.} [Planck Collaboration], Planck $2013$ results. XVI. Cosmological parameters. Astron.\ Astrophys.\  {\bf 571}, A16 (2014)
	
\bibitem{Aghanim:2018eyx} 
  N.~Aghanim {\it et al.} [Planck Collaboration], Planck $2018$ results. VI. Cosmological parameters. arXiv:1807.06209 [astro -ph.CO].

	
\bibitem{Copeland:2006wr} 
E.J.~Copeland, M.~Sami, S.~Tsujikawa, Dynamics of dark energy. Int.\ J.\ Mod.\ Phys.\ D {\bf 15}, 1753 (2006)
	
\bibitem{Frieman:2008sn1} 
  J.~Frieman, M.~Turner, D.~Huterer, Dark Energy and the Accelerating Universe.
  Ann.\ Rev.\ Astron.\ Astrophys.\ {\bf 46}, 385 (2008)
	
\bibitem{Nojiri:2017ncd} 
  S.~Nojiri, S.D.~Odintsov, V.K.~Oikonomou, Modified Gravity Theories on a Nutshell: Inflation, Bounce and Late-time Evolution. Phys.\ Rept.\  {\bf 692}, 1 (2017)
	
	
\bibitem{Zwicky:1933gu} 
F.~Zwicky, Die Rotverschiebung von extragalaktischen Nebeln. Helv.\ Phys.\ Acta {\bf 6}, 110 (1933)
[Gen.\ Rel.\ Grav.\ {\bf 41}, 207 (2009)]
	
	

\bibitem{AmendolaTsujikawa} L. Amendola, S. Tsujikawa, {\it Dark energy, theory and 
observations}, (Cambridge Univ. Press, Cambridge, England, 2010)

	\bibitem{FR-reviews7}
	S. Capozziello, V. Faraoni, {\it Beyond Einstein Gravity, Fundamental Theories of Physics},
Vol. 170 (Springer, Dordrecht, 2011)



\bibitem{Ratra:1987rm} 
  B.~Ratra, P.J.E.~Peebles, Cosmological Consequences of a Rolling Homogeneous Scalar Field.
  Phys.\ Rev.\ D {\bf 37}, 3406 (1988)


\bibitem{Copeland:1997et} 
  E.J.~Copeland, A.R.~Liddle, D.~Wands, Exponential potentials and cosmological scaling solutions.
  Phys.\ Rev.\ D {\bf 57}, 4686 (1998)


\bibitem{Caldwell:1997ii} 
  R.R.~Caldwell, R.~Dave, P.J.~Steinhardt, Cosmological imprint of an energy component with general equation of state. Phys.\ Rev.\ Lett.\  {\bf 80}, 1582 (1998)

\bibitem{Barreiro:1999zs} 
  T.~Barreiro, E.J.~Copeland, N.~J.~Nunes, Quintessence arising from exponential potentials.
  Phys.\ Rev.\ D {\bf 61}, 127301 (2000)


\bibitem{Sen:2002nu} 
  A.~Sen, Rolling tachyon. JHEP {\bf 0204}, 048 (2002)
	
\bibitem{Sen:2002in} 
  A.~Sen, Tachyon matter. JHEP {\bf 0207}, 065 (2002)


\bibitem{Padmanabhan:2002cp} 
  T.~Padmanabhan, Accelerated expansion of the universe driven by tachyonic matter.
  Phys.\ Rev.\ D {\bf 66}, 021301 (2002)
	
\bibitem{Abramo:2003cp} 
  L.R.W.~Abramo, F.~Finelli, Cosmological dynamics of the tachyon with an inverse power-law potential. Phys.\ Lett.\ B {\bf 575}, 165 (2003)

\bibitem{Copeland:2004hq} 
  E.J.~Copeland, M.R.~Garousi, M.~Sami, S.~Tsujikawa, What is needed of a tachyon if it is to be the dark energy?. Phys.\ Rev.\ D {\bf 71}, 043003 (2005)
  
	
\bibitem{Chiba:1999ka} 
  T.~Chiba, T.~Okabe, M.~Yamaguchi, Kinetically driven quintessence.
  Phys.\ Rev.\ D {\bf 62}, 023511 (2000)


\bibitem{ArmendarizPicon:2000dh} 
  C.~Armendariz-Picon, V.F.~Mukhanov, P.J.~Steinhardt, A Dynamical solution to the problem of a small cosmological constant and late time cosmic acceleration.
  Phys.\ Rev.\ Lett.\  {\bf 85}, 4438 (2000)


\bibitem{ArmendarizPicon:2000ah} 
  C.~Armendariz-Picon, V.F.~Mukhanov, P.J.~Steinhardt, Essentials of k essence.
  Phys.\ Rev.\ D {\bf 63}, 103510 (2001)


\bibitem{Piazza:2004df} 
  F.~Piazza, S.~Tsujikawa, Dilatonic ghost condensate as dark energy.
  JCAP {\bf 0407}, 004 (2004)

\bibitem{Gasperini:2001pc} 
  M.~Gasperini, F.~Piazza, G.~Veneziano, Quintessence as a runaway dilaton.
  Phys.\ Rev.\ D {\bf 65}, 023508 (2002)

\bibitem{Szydlowski:2006pz} 
  M.~Szydlowski, A.~Kurek, Testing and selection cosmological models with dark energy.
  AIP Conf.\ Proc.\  {\bf 861}, 1031 (2006)



	\bibitem{FR-reviews1}   
S. Capozziello, M. Francaviglia, Extended Theories of Gravity and their Cosmological and Astrophysical Applications.
Gen. Rel. Grav. \textbf{40}, 357 (2007)

	\bibitem{FR-reviews2}
A. De Felice, S. Tsujikawa, $f(R)$ theories. 
Living Rev. Rel. \textbf{13}, 3 (2010)

	\bibitem{FR-reviews3}
T.P. Sotiriou, V. Faraoni, $f(R)$ Theories of Gravity. 
Rev. Mod. Phys \textbf{82}, 451 (2010)

	\bibitem{FR-reviews4}
S. Nojiri, S.D. Odintsov, Unified cosmic history in modified gravity: from $F(R)$ theory to Lorentz non-invariant models. Phys. Rep. \textbf{505}, 59 (2011)

	\bibitem{FR-reviews5}
G.J. Olmo, Palatini Approach to Modified Gravity: $f(R)$ Theories and Beyond. 
Int. J. Mod. Phys. D \textbf{20}, 413 (2011)

	\bibitem{FR-reviews6}
S. Capozziello, M. De Laurentis, Extended Theories of Gravity. Phys. Rep. \textbf{509},  167 (2011)

\bibitem{Einstein} A. Einstein, Sitzungsber. Preuss. Akad. Wiss. Phys. Math. Kl. (1928) p. 217; p. 224 (1928)


\bibitem{TranslationEinstein} A. Unzicker, T. Case, Translation of Einstein's attempt of a unified field theory with teleparallelism. arXiv:physics/0503046 


\bibitem{Early-papers1}
A. Einstein, Math. Ann. {\bf 102}, 685 (1930)

\bibitem{Early-papers2}
A. Einstein, Sitzungsber. Preuss. Akad. Wiss. Phys. Math. Kl. 401 (1930)

\bibitem{Early-papers3}
C. Pellegrini, J. Pleba\'nski, K. Dan. Vidensk. Selsk. Mat. Fys. Skr. {\bf 2}, 2 (1962)

\bibitem{Early-papers4}
C. M{\o}ller, K. Dan. Vidensk. Selsk. Mat. Fys. Skr. {\bf 89}, 13 (1978)

\bibitem{Early-papers5}
K. Hayashi, T. Nakano, Extended translation invariance and associated gauge fields. Prog. Theor. Phys. {\bf 38}, 491 (1967)

\bibitem{Early-papers6}
K. Hayashi, T. Shirafuji, New General Relativity. Phys. Rev. D {\bf 19}, 3524 (1979);
Addendum: Phys.Rev. D {\bf 24}, 3312 (1982)


\bibitem{Aldrovandi-Pereira-book}  R. Aldrovandi, J.G. Pereira,
{\textit{Teleparallel Gravity: An Introduction}} (Springer, Dordrecht, 2013)


\bibitem{JGPereira2} J. G. Pereira, {\it Teleparallelism: a new insight into gravitation}, in {\it Springer Handbook of Spacetime}, ed. by A. Ashtekar and V. Petkov (Springer, Dordrecht, 2013), arXiv:1302.6983.


\bibitem{AndradeGuillenPereira-00} V.C. de Andrade, L.C.T. Guillen, J.G. Pereira,
Gravitational energy momentum density in teleparallel gravity. Phys. Rev. Lett. \textbf{84}, 4533 (2000)

\bibitem{Arcos:2005ec} 
  H.I.~Arcos, J.G.~Pereira, Torsion gravity: A Reappraisal.
  Int.\ J.\ Mod.\ Phys.\ D {\bf 13}, 2193 (2004)
	
	
\bibitem{Bengochea:2008gz} 
  G.R.~Bengochea, R.~Ferraro, Dark torsion as the cosmic speed-up.
  Phys.\ Rev.\ D {\bf 79}, 124019 (2009)


\bibitem{Linder:2010py} E.V. Linder, Einstein's Other Gravity and the Acceleration of the Universe. Phys. Rev. D \textbf{81}, 127301 (2010);
Erratum:[Phys. Rev. D \textbf{82}, 109902 (2010)]


\bibitem{Ferraro:2006jd}
  R.~Ferraro, F.~Fiorini, Modified teleparallel gravity: Inflation without inflaton.
  Phys.\ Rev.\ D {\bf 75}, 084031 (2007)





\bibitem{Iorio-Saridakis-2012} 	L. Iorio, E.N. Saridakis, Solar system constraints on $f(T)$ gravity. Mon.\ Not.\ Roy.\ Astron.\ Soc.\
\textbf{427},  1555 (2012)  

\bibitem{Iorio-2015}L. Iorio, N. Radicella, M.L.~Ruggiero, Constraining $f(T)$ gravity in the Solar System. JCAP \textbf{1508}, 021 (2015)

\bibitem{Farrugia-2016} G. Farrugia, J.L. Said, M.L.~Ruggiero, Solar System tests in $f(T)$ gravity. Phys.\ Rev.\ D \textbf{93}, 104034 (2016)

\bibitem{Bengochea-2011} G.R. Bengochea, Observational information for $f(T)$ theories and Dark Torsion. Phys. Lett. B \textbf{695}, 405 (2011)

\bibitem{Wei-Ma-Qi-2011} H. Wei, X.P. Ma, H.Y. Qi, $f(T)$ Theories and Varying Fine Structure Constant. Phys. Lett. B \textbf{703}, 74 (2011)

\bibitem{Capozziello-Luongo-Saridakis-2015} S.~Capozziello, O.~Luongo, E.N.~Saridakis, Transition redshift in $f(T)$ cosmology and observational constraints. Phys.\ Rev.\ D {\bf 91}, 124037 (2015)

\bibitem{Oikonomou-Saridakis-2016} V.K.~Oikonomou, E.N.~Saridakis, $f(T)$ gravitational baryogenesis. Phys.\ Rev.\ D \textbf{94}, 124005 (2016)

\bibitem{Nunes-Pan-Saridakis-2016} R.C.~Nunes, S.~Pan, E.N.~Saridakis, New observational constraints on $f(T)$ gravity from cosmic chronometers. JCAP \textbf{1608}, 011 (2016)

\bibitem{Wu-Yu-b-2010} P. Wu, H. Yu, The dynamical behavior of $f(T)$ theory. Phys. Lett. B \textbf{692}, 176 (2010)

\bibitem{Chen:2010va} 
  S.H.~Chen, J.B.~Dent, S.~Dutta and E.N.~Saridakis, Cosmological perturbations in $f(T)$ gravity. Phys.\ Rev.\ D {\bf 83}, 023508 (2011)


\bibitem{Dent-Duta-Saridakis-2011} J.B. Dent, S. Dutta, E.N. Saridakis, $f(T)$ gravity mimicking dynamical dark energy. Background and perturbation analysis. JCAP \textbf{1101},   009 (2011)

\bibitem{Zheng-Huang-2011} R. Zheng, Q.G. Huang, Growth factor in $f(T)$ gravity. JCAP \textbf{1103}, 002 (2011)

\bibitem{Wu:2012hs} 
  Y.P.~Wu, C.Q.~Geng, Matter Density Perturbations in Modified Teleparallel Theories. JHEP {\bf 1211}, 142 (2012)

\bibitem{Izumi:2012qj} K. Izumi, Y.C. Ong, Cosmological Perturbation in $f(T)$ Gravity Revisited. JCAP \textbf{1306}, 029 (2013)

\bibitem{Li:2011wu} B. Li, T.P. Sotiriou, J.D. Barrow, Large-scale Structure in $f(T)$ Gravity. Phys.\ Rev.\ D \textbf{83}, 104017 (2011)

\bibitem{Cai:2015emx} 
  Y.F.~Cai, S.~Capozziello, M.~De Laurentis, E.N.~Saridakis,
  ``f(T) teleparallel gravity and cosmology,
  Rept.\ Prog.\ Phys.\  {\bf 79}, 106901 (2016)
  
\bibitem{Nesseris:2013jea} 
  S.~Nesseris, S.~Basilakos, E.N.~Saridakis, L.~Perivolaropoulos, Viable $f(T)$ models are practically indistinguishable from $\Lambda$CDM.
  Phys.\ Rev.\ D {\bf 88}, 103010 (2013).

\bibitem{Basilakos:2016xob} S. Basilakos, Linear growth in power law $f(T)$ gravity. Phys.\ Rev.\ D \textbf{93}, 083007 (2016)

\bibitem{Wang-2011} T. Wang, Static solutions with spherical symmetry in $f(T)$ theories. Phys. Rev. D \textbf{84}, 024042 (2011)

\bibitem{Atazadeh:2012am} K. Atazadeh, M. Mousavi, Vacuum spherically symmetric solutions in $f(T)$ gravity. Eur.\ Phys.\ J.\ C \textbf{73}, 2272 (2013)

\bibitem{Ruggiero:2015oka} M.L. Ruggiero, N. Radicella, Weak-Field Spherically Symmetric Solutions in $f(T)$ gravity. Phys.\ Rev.\ D \textbf{91}, 104014 (2015)

\bibitem{Stars-in-f(T)} C.G. B\"ohmer, A. Mussa, N. Tamanini, Existence of relativistic stars in $f(T)$ gravity. Class.\ Quant.\ Grav.\ \textbf{28}, 245020 (2011)

\bibitem{Cosmography-2011} S. Capozziello, V.F. Cardone, H. Farajollahi, A. Ravanpak, Cosmography in $f(T)$-gravity. Phys. Rev. D \textbf{84}, 043527 (2011)

\bibitem{Liu-Reboucas-2012} D. Liu, M.J. Rebou\c{c}as, Energy conditions bounds on $f(T)$ gravity. Phys.\ Rev.\ D \textbf{86}, 083515 (2012)

\bibitem{Liu:2012kka} D. Liu, P. Wu, H. Yu, G\"odel-type universes in $f(T)$ gravity. Int.\ J.\ Mod.\ Phys.\ D \textbf{21}, 1250074 (2012)



\bibitem{Otalora:2017qqc} 
  G.~Otalora, M.J.~Rebou\c{c}as, Violation of causality in $f(T)$ gravity. Eur.\ Phys.\ J.\ C {\bf 77}, 799 (2017)
	
\bibitem{Cai:2018rzd} 
  Y.F.~Cai, C.~Li, E.N.~Saridakis, L.~Xue, $f(T)$ gravity after GW$170817$ and GRB$170817$A.
  Phys.\ Rev.\ D {\bf 97}, 103513 (2018)


\bibitem{Li:2018ixg} 
  C.~Li, Y.~Cai, Y.F.~Cai and E.N.~Saridakis, The effective field theory approach of teleparallel gravity, $f(T)$ gravity and beyond, arXiv:1803.09818 [gr-qc].




\bibitem{Li:2010cg} 
  B.~Li, T.P.~Sotiriou, J.D.~Barrow, $f(T)$ gravity and local Lorentz invariance. Phys.\ Rev.\ D {\bf 83}, 064035 (2011)


\bibitem{Krssak:2015oua}  M.~Kr\v{s}\v{s}\'ak, E.N.~Saridakis, The covariant formulation of $f(T)$ gravity. Class.\ Quant.\ Gravit.\ \textbf{33}, 115009 (2016)   


\bibitem{Harko:2014aja} 
T.~Harko, F.S.N.~Lobo, G.~Otalora, E.N.~Saridakis, $f(T,\mathcal{T})$ gravity and cosmology.
JCAP {\bf 1412}, 021 (2014)


\bibitem{Farrugia:2016pjh} 
G.~Farrugia, J.L.~Said, Growth factor in $f(T,\mathcal{T})$ gravity.
Phys.\ Rev.\ D {\bf 94}, 124004 (2016)
  


\bibitem{Harko:2014sja} 
  T.~Harko, F.S.N.~Lobo, G.~Otalora, E.N.~Saridakis, non-minimal torsion-matter coupling extension of $f(T)$ gravity. Phys.\ Rev.\ D {\bf 89}, 124036 (2014)


\bibitem{Carloni:2015lsa} 
  S.~Carloni, F.S.N.~Lobo, G.~Otalora, E.N.~Saridakis, Dynamical system analysis for a non-minimal torsion-matter coupled gravity. Phys.\ Rev.\ D {\bf 93}, 024034 (2016)
	
\bibitem{Nojiri:2004bi} 
  S.~Nojiri, S.D.~Odintsov, Gravity assisted dark energy dominance and cosmic acceleration.
  Phys.\ Lett.\ B {\bf 599}, 137 (2004)
  
\bibitem{Allemandi:2005qs} 
  G.~Allemandi, A.~Borowiec, M.~Francaviglia, S.D.~Odintsov, Dark energy dominance and cosmic acceleration in first order formalism.
  Phys.\ Rev.\ D {\bf 72}, 063505 (2005)

 
\bibitem{Nojiri:2006ri} 
  S.~Nojiri and S.D.~Odintsov, Introduction to modified gravity and gravitational alternative for dark energy. eConf C {\bf 0602061}, 06 (2006)
  [Int.\ J.\ Geom.\ Meth.\ Mod.\ Phys.\  {\bf 4}, 115 (2007)] 
 
  
\bibitem{Bertolami:2007gv} 
  O.~Bertolami, C.G.~Boehmer, T.~Harko, F.S.N.~Lobo, Extra force in $f(R)$ modified theories of gravity. Phys.\ Rev.\ D {\bf 75}, 104016 (2007)
	
\bibitem{Harko:2008qz} 
  T.~Harko, Modified gravity with arbitrary coupling between matter and geometry.
  Phys.\ Lett.\ B {\bf 669}, 376 (2008) 
	
\bibitem{Harko:2010mv} 
  T.~Harko, F.S.N.~Lobo, $f(R,L_{m})$ gravity.
  Eur.\ Phys.\ J.\ C {\bf 70}, 373 (2010) 
	
\bibitem{Bertolami:2009ic} 
  O.~Bertolami, J.~P\'aramos, Mimicking dark matter through a non-minimal gravitational coupling with matter. JCAP {\bf 1003}, 009 (2010) 
	

\bibitem{Bertolami:2013kca} 
  O.~Bertolami, P.~Fraz\~ao, J.~P\'aramos, Cosmological perturbations in theories with non-minimal coupling between curvature and matter.
  JCAP {\bf 1305}, 029 (2013)
	
\bibitem{Wang:2013fja} 
  J.~Wang, H.~Wang, Evolution of matter density perturbations in $f(R)$ theories of gravity with non-minimal coupling between matter and geometry. Phys.\ Lett.\ B {\bf 724}, 5 (2013)
	

\bibitem{Bertolami:2011fz} 
  O.~Bertolami, J.~P\'aramos, Mimicking the cosmological constant: Constant curvature spherical solutions in a non-minimally coupled model.
  Phys.\ Rev.\ D {\bf 84}, 064022 (2011) 

\bibitem{Bertolami:2010cw} 
  O.~Bertolami, P.~Fraz\~ao, J.~P\'aramos, Accelerated expansion from a non-minimal gravitational coupling to matter.
  Phys.\ Rev.\ D {\bf 81}, 104046 (2010) 

\bibitem{Gomes:2016cwj} 
  C.~Gomes, J.G.~Rosa, O.~Bertolami, Inflation in non-minimal matter-curvature coupling theories.
  JCAP {\bf 1706}, 021 (2017) 
	
\bibitem{Otalora:2018bso} 
  G.~Otalora, A.~\"{O}vg\"{u}n, J.~Saavedra and N.~Videla, Inflation from a nonlinear magnetic monopole field non-minimally coupled to curvature. JCAP {\bf 1806}, 003 (2018)

  
\bibitem{Birrell:1982ix} 
  N.D.~Birrell, P.C.W.~Davies, Quantum Fields in Curved Space (Cambridge
University Press, Cambridge, 1982)


\bibitem{Asselmeyer-Maluga:2016mvv} 
  T.Asselmeyer-Maluga, At the Frontier of Spacetime, Fundam.\ Theor.\ Phys.\  {\bf 183} (2016).


\bibitem{Groen:2007zz} 
  O.~Groen, S.~Hervik, Einstein's general theory of relativity: With modern applications in cosmology,
(New York, USA: Springer, 2007)	
 

\bibitem{Baker:2011wt} 
  T.~Baker, Phi Zeta Delta: Growth of Perturbations in Parametrized Gravity for an Einstein-de Sitter Universe. Phys.\ Rev.\ D {\bf 85}, 044020 (2012)

\bibitem{Wang:1998gt} 
  L.M.~Wang, P.J.~Steinhardt,
  Cluster abundance constraints on quintessence models.
  Astrophys.\ J.\  {\bf 508}, 483 (1998).
  
  \bibitem{Linder:2005in} 
  E.V.~Linder,
  Cosmic growth history and expansion history.
  Phys.\ Rev.\ D {\bf 72}, 043529 (2005).
  
\bibitem{Nesseris:2007pa} 
  S.~Nesseris, L.~Perivolaropoulos, Testing $\Lambda$CDM with the Growth Function $\delta(a)$: Current Constraints. Phys.\ Rev.\ D {\bf 77}, 023504 (2008)
  
\bibitem{Belloso:2011ms} 
A.~Bueno belloso, J.~Garcia-Bellido, D.~Sapone, A parametrization of the growth index of matter perturbations in various Dark Energy models and observational prospects using a Euclid-like survey. JCAP {\bf 1110}, 010 (2011)

\bibitem{Amendola:2016saw} 
  L.~Amendola {\it et al.}, Cosmology and fundamental physics with the Euclid satellite.
  Living Rev.\ Rel.\  {\bf 21}, 2 (2018)
	
	
\bibitem{Weinberg:2012es} 
  D.H.~Weinberg, M.J.~Mortonson, D.J.~Eisenstein, C.~Hirata, A.G.~Riess, E.~Rozo, 
	Observational Probes of Cosmic Acceleration. Phys.\ Rept.\  {\bf 530}, 87 (2013)
	

\bibitem{C_M_Will}
	C.M. Will, {\it Theory and Experiment in Gravitational Physics}, (Cambridge: Cambridge University Press, 1981)
 

\bibitem{Sotiriou:2008it} 
  T.P.~Sotiriou, V.~Faraoni, Modified gravity with $R$-matter couplings and (non-)geodesic motion.
  Class.\ Quant.\ Grav.\  {\bf 25}, 205002 (2008)


\bibitem{Sola:2017znb} 
  J.~Sol\`a, A.~G\'omez-Valent, J.~de Cruz P\'erez, The $H_0$ tension in light of vacuum dynamics in the Universe.
  Phys.\ Lett.\ B {\bf 774}, 317 (2017)


\bibitem{Kazantzidis:2018rnb} 
  L.~Kazantzidis, L.~Perivolaropoulos, Evolution of the $f\sigma_8$ tension with the Planck15/$\Lambda$CDM determination and implications for modified gravity theories. Phys.\ Rev.\ D {\bf 97}, 103503 (2018)


\bibitem{Albarran:2016mdu} 
  I.~Albarran, M.~Bouhmadi-L\'opez, J.~Morais, Cosmological perturbations in an effective and genuinely phantom dark energy Universe. Phys.\ Dark Univ.\  {\bf 16}, 94 (2017)
	

\bibitem{Gomez-Valent:2018nib} 
  A.~G\'omez-Valent, J.~Sol\`a Peracaula, Density perturbations for running vacuum: a successful approach to structure formation and to the $\sigma_8$-tension. Mon.\ Not.\ Roy.\ Astron.\ Soc.\  {\bf 478}, 126 (2018)

\bibitem{Abedi:2018lkr} 
  H.~Abedi, S.~Capozziello, R.~D'Agostino, O.~Luongo, Effective gravitational coupling in modified teleparallel theories.
 Phys.\ Rev.\ D {\bf 97}, 084008 (2018)




	
	



\end{thebibliography}
\end{document}